\documentclass[aps,reprint,nofootinbib,twocolumn,superscriptaddress,longbibliography]{revtex4-2}
\usepackage[T1]{fontenc}
\usepackage{bbold}
\usepackage{physics}
\usepackage{comment}
\usepackage{natbib}
\usepackage{slashed}
\usepackage[english]{babel}
\usepackage{graphicx,epstopdf}
\usepackage{amssymb,amsmath}
\usepackage{dcolumn}
\usepackage{bm}
\usepackage{color}
\usepackage{enumerate}
\usepackage[colorlinks=true,citecolor=blue]{hyperref}
\hypersetup{colorlinks=true,citecolor=blue,linkcolor=blue,urlcolor=blue}
\usepackage{soul}
\usepackage{tabularx}
\usepackage[dvipsnames]{xcolor}
\usepackage{cancel}

\begin{document}

\preprint{APS/123-QED}

\title{Theory of off-diagonal disorder in multilayer topological insulator}

\author{Z.Z. Alisultanov}
\email{zaur0102@gmail.com}
\affiliation{Abrikosov Center for Theoretical Physics, MIPT, Dolgoprudnyi, Moscow Region 141701, Russia}
\affiliation{Institute of Physics of DFRS, Russian Academy of Sciences, Makhachkala, 367015, Russia}
\author{A. Kudlis}
\email{andrewkudlis@gmail.com}
\affiliation{Science Institute, University of Iceland, Dunhagi 3, IS-107, Reykjavik, Iceland}
\affiliation{Abrikosov Center for Theoretical Physics, MIPT, Dolgoprudnyi, Moscow Region 141701, Russia}

\begin{abstract}
 We study multilayer topological insulators with random interlayer tunneling, known as off-diagonal disorder. Within the Burkov-Balents model a single Hermitian defect creates a bound state whose energy crosses the middle of the gap in the trivial phase but never in the topological phase; a non-Hermitian defect splits this level yet preserves the same crossing rule, so the effect serves as a local marker of topology. However, the key distinction persists: the bound state crosses zero in the trivial phase but not in the topological phase. Two complementary diagrammatic approaches give matching densities of states for the normal, topological, Weyl and anomalous quantum Hall regimes. Off diagonal disorder inserts bulk states into the gap and can close it: the Weyl phase remains robust under strong disorder, whereas the anomalous quantum Hall phase survives only for weak fluctuations, and the added bulk states shrink the Hall plateau, clarifying experimental deviations. Finally, we analyze edge modes. Uniform disorder shortens their localization length slightly, while Gaussian and Lorentzian disorder enlarge it and in the Gaussian case can even delocalize the edges. Although chirality is maintained, the enhanced overlap permits tunneling between opposite edges and pulls the longitudinal conductance away from its quantized value. 
\end{abstract}

\maketitle

\section{Introduction} \label{introduction}

The influence of disorder on systems with non-trivial topology remains one of the central issues in the field~\cite{klitzing1980new,qi2011topological,liu2025interplay,zhang2009experimental,chen2010gate,lu2011competition,jiang2012stabilizing,jiang2014transport,shi2021disorder,alisultanov2024disorder,wu2016disorder}. Non-trivial band topology markedly affects several analytical features that arise in the presence of disorder. For example, impurity-induced bound states inside the gap and the associated zeros of the Green’s function have been proposed as local detectors of a system’s topology~\cite{Slager,Diop}. In addition to \emph{diagonal} disorder, where the on-site energy fluctuates, there exists \emph{off-diagonal} disorder, in which the inter-site hopping parameters are random.  The properties of this type of disorder were first examined by Dyson, who obtained several exact results~\cite{Dyson}.  Important subsequent contributions include Refs.~\cite{soukoulis1981off,theodorou1976extended,brouwer2000density,cheraghchi2005localization,cheraghchi2006scaling,saha2023localization}.  In Refs.~\cite{Raghavan,Harris,Timothy_Ziman}, localization theories for special models showed that off-diagonal disorder can produce anomalous localization. Ref.~\cite{Inui} explored further unusual features of in-gap states for this kind of disorder in 1D system.

\begin{figure}[t]
    \centering
    \includegraphics[width=1\linewidth]{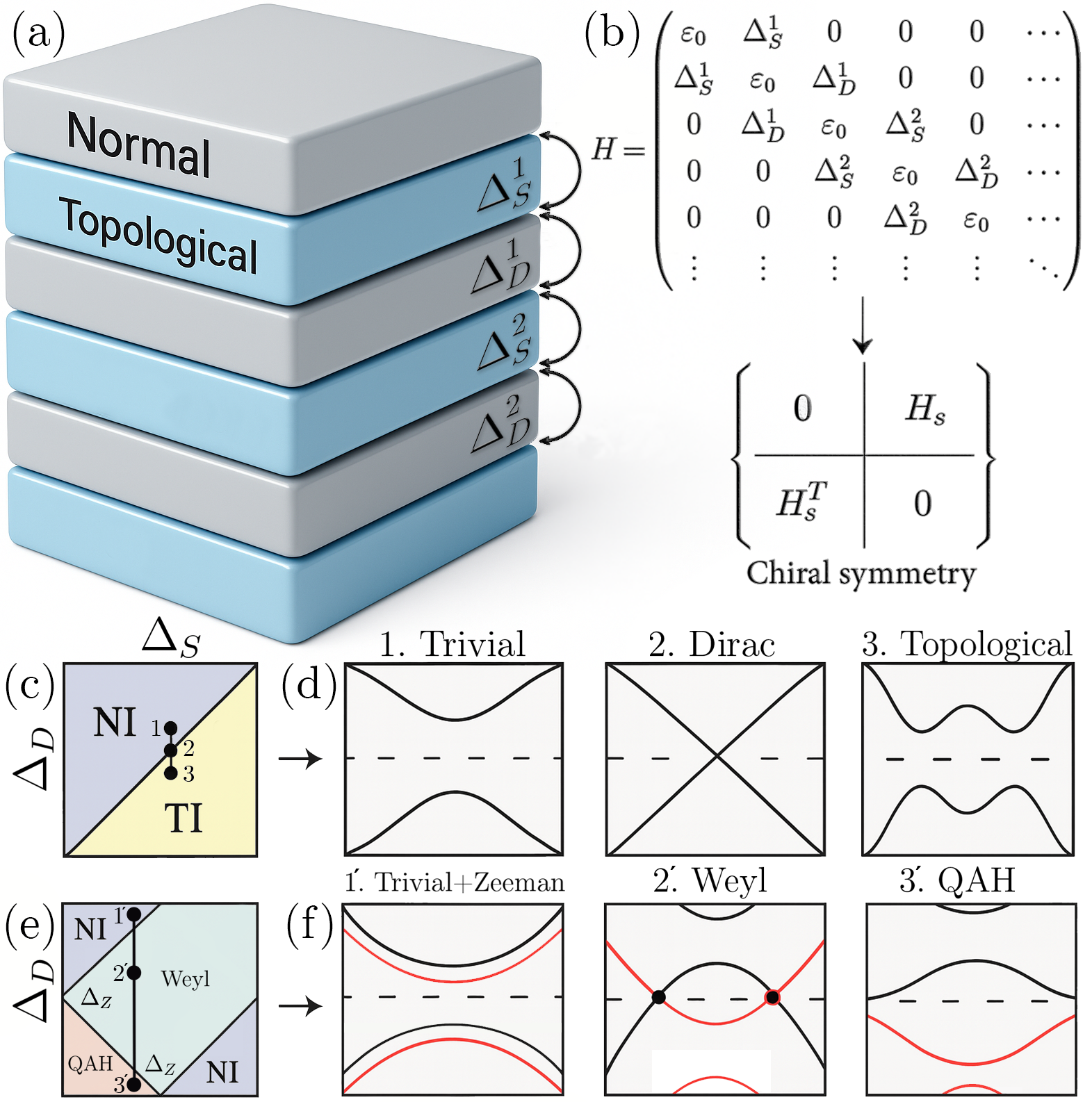}
    \caption{(a) Upper-left panel shows the three–dimensional sketch of the van‑der‑Waals stack studied here:  
blue slabs are topological‑insulator (TI) layers, grey slabs are trivial (normal‑insulator, NI) spacers. Random interlayer couplings are indicated by the symbols
$\Delta_{S}^{1,2,\dots}$ (tunneling between topological surface states through a TI film) and $\Delta_{D}^{1,2,\dots}$ (tunneling between topological surface states across a NI layer). (b) Upper-right panel demonstrates the tight‑binding matrix of the Burkov–Balents Hamiltonian in the layer basis. Reordering the basis exposes its chiral (off‑block‑diagonal) structure, so every eigenstate has a partner at the opposite energy. \textbf{(c)}~Phase diagram in the clean limit, plotted in the $(\Delta_{S},\Delta_{D})$ plane.  
The line $\Delta_{S}=\Delta_{D}$ separates a trivial insulator (NI) from a topological insulator (TI); points~1, 2, 3 mark parameter sets whose bulk dispersions are shown in~(d). (d)~Corresponding band structures along $k_{z}$: (1) trivial gap, (2) Dirac point at $\Delta_{S}=\Delta_{D}$, (3) topological gap. (e)~Phase diagram after including a Zeeman splitting~$\Delta_{Z}$. Besides the NI regions it now hosts a Weyl semimetal (Weyl) and an anomalous quantum Hall phase (QAH). Points~1$'$, 2$'$, 3$'$ correspond to the bulk spectra in~(f): (1$'$) Zeeman‑split trivial insulator, (2$'$) Weyl phase with two nodes (red), (3$'$) gapped QAH phase with chiral edge modes (red dispersions). Dashed horizontal lines denote the mid‑gap energy throughout.}
\label{fig:skt}
\end{figure}

The impact of disorder on topological systems is of particular interest for several reasons.  In experiments, such disorder may account for deviations of chiral‐channel transport from the universal values set by fundamental constants~\cite{konig2007quantum}.  The effect is especially pronounced in systems that exhibit bulk‐inversion asymmetry (BIA) or structural‐inversion asymmetry (SIA).  These asymmetries shift channels with different spin projections, so counter-propagating edge modes no longer possess a well-defined spin polarization~\cite{rothe2010fingerprint,maciejko2010magnetoconductance}.  For elastic scattering this leaves chiral transport intact, as it remains protected by time-reversal symmetry.  Under inelastic scattering, however, a finite overlap can arise between states of different momenta, which may affect, for instance, the edge conductance. Studying other possible sources of the observed departures from ideal behavior remains a major objective.

Disorder-induced topological phases in two‐ and three-dimensional systems -- dubbed 2D and 3D \emph{Anderson topological insulators} -- are of special interest~\cite{Liu2009,Groth,Guo}.  In such materials disorder simultaneously causes Anderson localization and renormalizes the topological mass, even inverting its sign and driving the system into a non-trivial phase~\cite{Groth}.  Hence disorder can generate chiral edge states in systems that are topologically trivial in the clean limit.  A recent experiment reports a \emph{topological Anderson Chern insulator} in two-dimensional MnBi$_4$Te$_7$~\cite{Wang25}, further stimulating research in this direction.

Van der Waals multilayers that alternate thin topological‐insulator (TI) and trivial‐insulator sheets form a convenient platform for realizing exotic topological phases~\cite{Fan,Kandala,Hesjedal,Chong,Eremeev,Liu,Zaur,belopolski2025synthesis,tang2025quantum}.  
A key advantage is that magnetic impurities can be introduced into the trivial layers, producing the required band splitting through proximity effects without disrupting the intrinsic TI properties.  Such heterostructures reveal subtle signatures of the quantum metric, including an anomalous Hall response~\cite{gao2023quantum}.  
They also enable a more robust quantum spin Hall (QSH) phase at elevated temperatures -- for example in WTe$_2$ -- whereas conventional platforms such as HgTe/CdTe quantum wells exhibit QSH behavior only at cryogenic temperatures, limiting their utility.

\allowdisplaybreaks

In a recent study~\cite{Alisultanov_multilayer} we analyzed how non-magnetic disorder affects the electronic states of a multilayer topological insulator.  Considering both the presence and absence of Zeeman splitting, we showed that such disorder can drive transitions between the system’s topological phases.  That work also initiated an examination of off-diagonal disorder: a Green-function expansion in localized states revealed that it can markedly reshape the density of states, narrowing -- and even closing -- the gap.  

In this work we present a detailed analysis of off-diagonal disorder in multilayer topological insulators. We treat both a single off-diagonal defect and a finite concentration of such defects, comparing results from two independent approaches.  We also study how this disorder influences topological transport in the topological-insulator and anomalous quantum Hall regimes. Fig.~\ref{fig:skt} summarizes the physical set‑up analyzed in this work. It shows the alternating TI/NI van-der-Waals stack, the corresponding block-structured Burkov–Balents Hamiltonian with its chiral form, and the clean‑limit as well as Zeeman‑modified phase diagrams that motivate the parameter regimes explored below.  
This sketch will serve as a visual guide for the calculations presented in the subsequent sections.

Following this introduction, the remainder of the paper is organized as follows.  In Section~II, we extend the Burkov–Balents Hamiltonian to include off-diagonal disorder and analyze its symmetry properties.  
In Section~III, we develop the Green-function formalism using two complementary approaches: expansions in localized and Bloch states.  Section~IV addresses the single-defect problem for off-diagonal disorder, treating both Hermitian and non-Hermitian cases. Section~V presents numerical results for the density of states alongside selected analytical calculations. This section also examines the localization length of edge modes and evaluates the correction to edge conductance arising from divergence of the localization length. Finally, the main conclusions are summarized in the Conclusion.  Supplementary calculations appear in the Appendices.

\section{Burkov-Balents model with off-diagonal disorder}

We now formulate the problem in the presence of off-diagonal disorder. In its most general form the Burkov–Balents Hamiltonian~\cite{Burkov} with fluctuating tunneling amplitudes is
\begin{multline}
\mathcal{H}=%
\sum_{\mathbf{k}_{\perp},i}
c_{\mathbf{k}_{\perp}i}^{\dagger}
\Bigl[\,\tau^{z}\upsilon_{F}\!\left(\hat{\bm{z}}\!\times\!\bm{\sigma}\right)\!\cdot\!\mathbf{k}_{\perp}
+\Delta_{S}^{i}\tau^{x}\Bigr]
c_{\mathbf{k}_{\perp}i}\\
+\sum_{\mathbf{k}_{\perp},i,j}
c_{\mathbf{k}_{\perp}i}^{\dagger}
\Bigl[\Delta_{+}^{i}\tau^{+}\delta_{j,i+1}
+\Delta_{-}^{i}\tau^{-}\delta_{j,i-1}\Bigr]
c_{\mathbf{k}_{\perp}j},
\label{off-diagonal Hamiltonian}
\end{multline}
where $\Delta_{S}^{i}$ and $\Delta_{\pm}^{i}$ are random functions of the layer index~$i$.  
We write $\Delta_{S}^{i}=\Delta_{S}+\eta_{S}^{i}$ and $\Delta_{\pm}^{i}=\Delta_{D}+\eta_{\pm}^{i}$,  
with $\Delta_{S}$ and $\Delta_{D}$ the uniform (regular) parts and $\eta_{S}^{i},\eta_{\pm}^{i}$ the random fluctuations. We adopt the simpler case in which only $\Delta_{S}$ fluctuates. We will use this simplification until Section V, where we will consider general case in which both tunneling parameters fluctuate.  
These fluctuations act much like a random on-site potential-- because they enter with the Kronecker factor $\delta_{ij}$ -- even though $\Delta_{S}$ itself is an off-diagonal parameter that couples edge modes via the Pauli matrix $\tau^{x}$.  Neglecting fluctuations in $\Delta_{D}$ streamlines the analysis without obscuring the essential physics.

In the absence of disorder, the momentum-space Hamiltonian reads
\begin{multline}
\mathcal{H}_{\mathbf{k}}^{0}
=\upsilon_{F}\tau^{z}\otimes\left(\hat{\bm{z}}\times\bm{\sigma}\right)\!\cdot\!\mathbf{k}_{\perp}
+\tau^{x}\otimes\sigma_{0}\left(\Delta_{S}\right.\\-\left.\Delta_{D}\cos k_{z}d\right)
+\tau^{y}\otimes\sigma_{0}\,\Delta_{D}\sin k_{z}d .
\end{multline}
This Hamiltonian is time-reversal symmetric: 
$\mathcal{T}\mathcal{H}_{-\mathbf{k}}\mathcal{T}^{-1}=\mathcal{H}_{\mathbf{k}}$ 
with $\mathcal{T}=i\tau^{0}\otimes\sigma_{y}K$.  
The gapless (Dirac semimetal) phase occurs when $\Delta_{S}=\Delta_{D}$, giving a Dirac point at $\mathbf{k}_{\perp}=0$, $k_{z}=0$.  
In this regime the system also enjoys a mirror symmetry with respect to the plane that passes through the interface between the topological and trivial layers.  
Under this reflection $k_{z}\!\to\!-k_{z}$ and $\sigma_{x,y}\!\to\!-\sigma_{x,y}$.  
The corresponding operator is $\mathcal{R}=\tau^{x}\otimes\sigma_{z}$, and in the gapless phase it satisfies $\mathcal{R}\mathcal{H}_{-k_{z}}\mathcal{R}^{-1}=\mathcal{H}_{k_{z}}$. 

The gapped phase $\Delta_{S}\neq\Delta_{D}$ is likewise time-reversal invariant: the off-diagonal mass term 
$\tau^{x}\!\otimes\!\sigma_{0}\,(\Delta_{S}-\Delta_{D})$ emerges without breaking $\mathcal{T}$ symmetry,  
but it \emph{does} break the mirror symmetry $\mathcal{R}$. By contrast, a diagonal mass term of the form $\Delta\,\tau^{z}\!\otimes\!\sigma_{z}$ cannot arise, since it violates both $\mathcal{T}$ symmetry and chiral symmetry. Within the gapped regime the multilayer structure can adopt two distinct states:  
a \emph{normal insulator} for $\Delta_{S}>\Delta_{D}$ and a \emph{topological insulator} for $\Delta_{S}<\Delta_{D}$.   This distinction cannot be attributed to the symmetries of the bulk Hamiltonian $\mathcal{H}_{\mathbf{k}}^{0}$ alone. Instead, it is captured by a Fu–Kane $\mathbb{Z}_{2}$ invariant~$v$, defined via  
$(-1)^{v}=\operatorname{sgn}(\Delta_{S}-\Delta_{D})$.  
Alternatively, one may introduce a winding number analogous to that in the SSH model. These symmetry considerations place the resulting phases within the Altland–Zirnbauer classification scheme~\cite{chiu2016classification}.

When off-diagonal disorder is present, the topological invariants mentioned above cease to provide useful information.   Nevertheless, the Hamiltonian retains a chiral symmetry, $\mathcal{K}\mathcal{H}\mathcal{K}^{-1}=-\mathcal{H}$, for example with $\mathcal{K}=\tau^{z}\!\otimes\!\sigma_{z}$.  
Consequently, every solution of the wave equation is doubled:  if $\lvert\psi\rangle$ is an eigenstate, then $\mathcal{K}\lvert\psi\rangle$ is also an eigenstate with the opposite energy.

We can exploit chiral symmetry to analyze zero modes.  
Writing the wave equation for such modes yields the following iterative relation for the wave-function amplitude on the $(2n\!+\!1)$-st layer:
\begin{gather}
\psi_{2n+1}=(-1)^{n}\frac{\Delta_{S}^{1}}{\Delta_{D}}\frac{\Delta_{S}^{2}}{\Delta_{D}}\dots\frac{\Delta_{S}^{n}}{\Delta_{D}}\psi_{1}.
\end{gather}
Because the $\Delta_{S}^{i}$ are random, a first rough estimate of the edge-mode localization length -- here identical to the penetration depth -- can be obtained from~\cite{soukoulis1981off}
\begin{gather}
\frac{1}{L(\epsilon=0)}=-\lim_{n\to\infty}\frac{1}{n}\ln\!\left|\frac{\psi_{2n+1}}{\psi_{1}}\right|.
\label{localization lenght}
\end{gather}
Without disorder this gives
\begin{gather}
\frac{1}{L(\epsilon=0)}=-\ln\!\left|\frac{\Delta_{S}}{\Delta_{D}}\right|.
\end{gather}
Thus $L\!\to\!\infty$ when $\Delta_{S}=\Delta_{D}$, so the edge state becomes bulk-like and the system enters the Dirac semimetal phase.  
For $\Delta_{S}>\Delta_{D}$ one finds $L(\epsilon=0)<0$, which is unphysical -- no edge state exists and the system is a trivial insulator.  
When $\Delta_{S}<\Delta_{D}$ the localization length is finite and positive (tending to zero as $\Delta_{S}\!\to\!0$), corresponding to an edge state in the topological phase.

With disorder, the quantities $\ln|\Delta_{S}^{k}/\Delta_{D}|$ are random.  
Averaging the sum in~\eqref{localization lenght} and invoking the central-limit theorem yields $1/L(\epsilon=0)=0$; off-diagonal disorder therefore delocalizes the edge state.
\begin{figure}[t!]
\centering
\includegraphics[width=1\columnwidth]{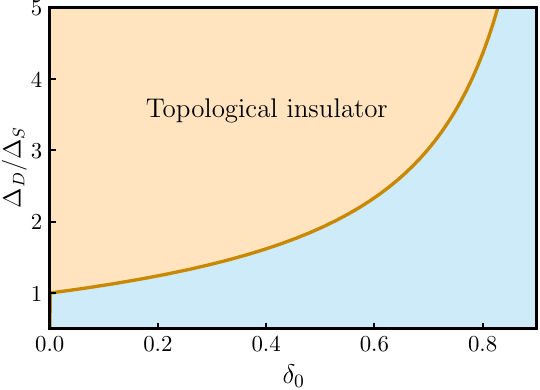}
\caption{Phase diagram in the $(\delta_{0},\,\Delta_{D}/\Delta_{S})$ plane, where  
$\delta_{0}=\eta_{0}/\Delta_{S}$ measures the amplitude of off–diagonal disorder and  
$\Delta_{D}/\Delta_{S}$ sets the clean‑limit band inversion.  The solid curve is the critical line obtained from Eq.~\eqref{localization condition}. Blue region (below the curve): the disorder satisfies the inequality in Eq.~(9), the edge state delocalizes, and the topological phase is destroyed.  
Orange region (above the curve): the inequality is violated, the edge mode remains localized, and the system retains its topological‑insulator character.%
}
\label{wave func}
\end{figure}
An alternative viewpoint employs the cumulant method for disordered systems~\cite{yonezawa1966note}.  
Introducing the mean wavefunction,
\begin{gather}
\overline{\psi}_{2k+1}
=\exp\!\bigl\langle \ln\psi_{2k+1}\bigr\rangle
=\Bigl(-\exp\!\bigl\langle \ln\frac{\Delta_{S}^{i}}{\Delta_{D}}\bigr\rangle\Bigr)^{k}\overline{\psi}_{1},
\end{gather}
and taking $\Delta_{S}^{i}=\Delta_{S}+\eta_{i}$ with Anderson averaging  
$\langle\cdots\rangle=(2\eta_{0})^{-1}\!\int_{-\eta_{0}}^{\eta_{0}}\!\cdots\,d\eta$, we find
\begin{gather}
\overline{\psi}_{2k+1}
=\Bigl(-\frac{\Delta_{S}}{\Delta_{D}}\exp\!\Bigl[\frac{1+\delta_{0}}{2\delta_{0}}
\ln\frac{1+\delta_{0}}{1-\delta_{0}}-1\Bigr]\Bigr)^{k}\overline{\psi}_{1},
\end{gather}
where $\delta_{0}=\eta_{0}/\Delta_{S}$.  
In the clean limit $\Delta_{S}^{i}=\Delta_{S}$, this reduces to
\begin{gather}
\overline{\psi}_{2k+1}=\Bigl(-\frac{\Delta_{S}}{\Delta_{D}}\Bigr)^{k}\psi_{1}.
\end{gather}
Hence for $\Delta_{S}<\Delta_{D}$ the edge-state amplitude decays exponentially away from the boundary.  
Off-diagonal disorder changes this: the state becomes extended when
\begin{gather}
\frac{1+\delta_{0}}{2\delta_{0}}\ln\frac{1+\delta_{0}}{1-\delta_{0}}-1>\ln\frac{\Delta_{S}}{\Delta_{D}},
\label{localization condition}
\end{gather}
which we denote the condition for destroying the topological phase.  
The corresponding phase diagram is shown in Fig.~\ref{wave func}.

A more detailed study of localization and delocalization is presented in Section~5.

\section{Green's function formalism}

In this section we briefly outline the formalism used in our analysis. Two complementary methods are employed to enable cross-checking of results.

We write the Hamiltonian as a sum of a regular part and a perturbation:
\begin{equation}
\mathcal{H} = \sum_{\mathbf{k}_{\perp},i,j}
c_{\mathbf{k}_{\perp}i}^{\dagger}
\left[ T_{ij}(\mathbf{k}_{\perp}) + V_i\,\delta_{i,j} \right]
c_{\mathbf{k}_{\perp}j},
\end{equation}
where
\begin{align}
T_{ij}(\mathbf{k}_{\perp}) &= 
\left(\tau^{z}\upsilon_{F}(\hat{\bm{z}}\!\times\!\bm{\sigma})\cdot\mathbf{k}_{\perp} + \Delta_{S}\tau^{x} \right)\delta_{i,j} \nonumber\\
&\quad + \tfrac{1}{2}\Delta_{D}\tau^{+}\delta_{j,i+1}
+ \tfrac{1}{2}\Delta_{D}\tau^{-}\delta_{j,i-1}, \\
V_i &= \eta_{S}^{i}\,\tau^{x}.
\end{align}
We now briefly comment on the effect of Zeeman splitting.  In its presence, the Hamiltonian becomes
\begin{equation}
\mathcal{H}_{\mathbf{k}} \rightarrow 
\mathcal{H}_{\mathbf{k}} + \tau^{0}\sigma_{z}\Delta_{Z},
\label{eqn:hami_zeem}
\end{equation}
where $\Delta_{Z}$ is the Zeeman splitting.  
This term leads to the emergence of Weyl nodes and induces a transition into the anomalous quantum Hall (AQH) phase (see~\cite{Burkov,Alisultanov_multilayer}).

\subsection{Expansion in localized states}

This approach was introduced in Ref.~\cite{Alisultanov_multilayer}, where full details are given; here we summarize the essential formulas.

Following Refs.~\cite{Alisultanov_multilayer,Toyozawa,Matsubara,J.M.Ziman}, we expand the Green function in the layer (node) representation with respect to the interlayer hopping
$t_{ij}=\tfrac{1}{2}\Delta_{D}\bigl(\tau^{+}\delta_{j,i+1}+\tau^{-}\delta_{j,i-1}\bigr)$:
\begin{multline}
G_{ij}\!\left(\mathbf{k}_{\perp},\epsilon\right)
=S_{i}\!\left(\mathbf{k}_{\perp}\right)\delta_{ij}
+S_{i}\!\left(\mathbf{k}_{\perp}\right)t_{ij}S_{j}\!\left(\mathbf{k}_{\perp}\right)\\
+\sum_{j'}S_{i}t_{ij'}S_{j'}t_{j'j}S_{j}+\dots ,
\end{multline}
with
\begin{equation}
S_{i}\!\left(\mathbf{k}_{\perp}\right)
=\bigl(\epsilon-\tau^{z}\upsilon_{F}(\hat{\bm{z}}\!\times\!\bm{\sigma})\!\cdot\!\mathbf{k}_{\perp}
-\Delta_{S}\tau^{x}-V_{i}\bigr)^{-1}.
\end{equation}
(See Appendix~A for the derivation.)  
Fourier transforming gives
\begin{multline}
G_{k_{z}k_{z}'}\!\left(\mathbf{k}_{\perp},\epsilon\right)
=\sigma_{k_{z}k_{z}'}
+\sum_{k_{z}''}\sigma_{k_{z}k_{z}''}t_{k_{z}''}\sigma_{k_{z}''k_{z}'}\\
+\sum_{k_{z}'',k_{z}'''}\!\sigma_{k_{z}k_{z}''}t_{k_{z}''}\sigma_{k_{z}''k_{z}'''}t_{k_{z}'''}\sigma_{k_{z}'''k_{z}'}+\dots ,
\label{Green_func_equation}
\end{multline}
where 
$\sigma_{k_{z}k_{z}'}=\tfrac{1}{N}\sum_{j}S_{j}
\exp\!\bigl[i(k_{z}-k_{z}')z_{j}\bigr]$ and  
$t_{k_{z}}=\Delta_{D}\bigl(\tau^{x}\cos k_{z}a-\tau^{y}\sin k_{z}a\bigr)$.

\begin{figure}
	\centering 
	\includegraphics[width=1\linewidth]{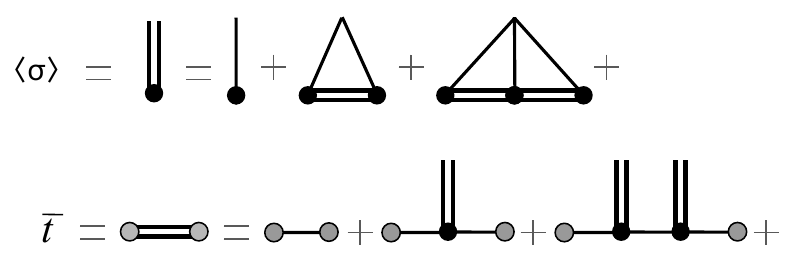}	
	\caption{Diagrammatic expansions for the quantities $\left\langle \sigma\right\rangle $ and $\overline{t}$. Gray circles in the lower diagrams indicate identical indices. Intermediate indices can be any, but differ from the edge indices (gray). This requirement simply corresponds to the exclusion of repeated counting of the same charts. These series correspond to the Eq.~\eqref{locator} and~\eqref{interactor}.} 
	\label{fig:Diagram series}%
\end{figure}

To account for disorder we must average the Green function.  
Introducing the renormalized locator,
\begin{gather}
\langle\sigma\rangle
=\bigl\langle\bigl(S_{m}^{-1}-\overline{t}\,\bigr)^{-1}\bigr\rangle,
\label{locator}
\end{gather}
the diagrammatic series in Fig.~\ref{fig:Diagram series} yields
\begin{equation}
\overline{t}\!\left(\mathbf{k}_{\perp}\right)
=\langle\sigma\rangle^{-1}\bigl(\langle G\rangle t\bigr)_{mm}
=\langle\sigma\rangle^{-1}\!
\sum_{k_{z}}\!\langle\overline{G}_{\mathbf{k}}\rangle t_{k_{z}},
\label{interactor}
\end{equation}
with
\begin{equation}
\langle\overline{G}_{\mathbf{k}}\rangle
=\bigl(\langle\sigma\rangle^{-1}
+\overline{t}\!\left(\mathbf{k}_{\perp}\right)-t_{k_{z}}\bigr)^{-1}.
\label{self_green_func}
\end{equation}
These self-consistent equations were first obtained in Refs.~\cite{SHIBA,Blackman}.  
In the clean limit $\langle\sigma\rangle=(S_{m}^{-1}-\overline{t})^{-1}$, recovering the unperturbed Green function.  
Equations~\eqref{locator}–\eqref{self_green_func}, averaged over a homogeneous ensemble (as in the Anderson model), constitute the \emph{self-consistent locator} method employed for the density-of-states calculations.

\subsection{Expansion in Bloch states}

For comparison we also perform calculations using a perturbative expansion of the Green function in the Bloch basis (BSE method; see, e.g., Ref.~\cite{J.M.Ziman}):
\begin{multline}
\bigl\langle G_{ij}\bigl(\mathbf{k}_{\perp},\epsilon\bigr)\bigr\rangle
=G_{ij}^{0}
+\sum_{l}G_{il}^{0}\,\langle V_{l}\rangle\,G_{lj}^{0}\\
+\sum_{l l'}G_{il}^{0}\,\langle V_{l}G_{l l'}^{0}V_{l'}\rangle\,G_{l'j}^{0}+\dots
\end{multline}
To simplify the averaging we adopt the $t$‑matrix approximation, rewriting the series as
\begin{multline}
\bigl\langle G_{ij}\bigl(\mathbf{k}_{\perp},\epsilon\bigr)\bigr\rangle
=G_{ij}^{0}
+\overline{t_{l}}\sum_{l}G_{il}^{0}G_{l j}^{0}\\
+\overline{t_{l}}^{2}\sum_{l\neq l'}G_{il}^{0}G_{l'j}^{0}+\dots
\end{multline}
where $t_{l}=(1-V_{l}G_{ll}^{0})^{-1}V_{l}$ and $\overline{t_{l}}=\langle t_{l}\rangle$.  
Because repeating indices are absent, the series cannot be summed directly to a Dyson equation.  
We therefore employ the \emph{$t$‑matrix cancellation trick} of Ref.~\cite{J.M.Ziman}: replace
$
\overline{t_{l}}\to\overline{\upsilon}
=\overline{t_{l}}\bigl(1+G_{ll}^{0}\overline{t}\bigr)^{-1},
$
with $G_{ll}^{0}=\sum_{k_{z}}G_{\mathbf{k}}^{0}$, and simultaneously restore the repeated indices.  
The series then becomes summable, yielding a Dyson equation that is solved by Fourier transformation,
\begin{gather}
\bigl\langle G_{k}\bigr\rangle
=\Bigl[\bigl(G_{k}^{0}\bigr)^{-1}-\overline{\upsilon}\Bigr]^{-1}
=\Bigl[\epsilon-\overline{\upsilon}-T_{k_{z}}\bigl(\mathbf{k}_{\perp}\bigr)\Bigr]^{-1}.
\label{GF t-matrix}
\end{gather}

Because the $t$‑matrix cancellation is applied \emph{after} disorder averaging, the resulting series retains disorder information.  
One can show that this trick corresponds to summing RPA‑type diagrams~\cite{aiyer1969pair}; accordingly, we refer to the approximation as \textit{t‑matrix+RPA}.  
To make it self‑consistent, we replace $G_{ll}^{0}\to\langle G_{ll}\rangle=\sum_{k_{z}}\langle G_{k}\rangle$ in $\overline{\upsilon}$.

Analytical expressions for $G_{ll}^{0}$, needed both for the t‑matrix+RPA scheme and for the single‑defect problem, are collected in Appendix~B.

\section{Problem of a single off‑diagonal defect}

\subsection{Hermitian case}

We begin with a single off‑diagonal defect in an MTI. Analysing one defect often reveals the general impact of disorder, and this problem is exactly solvable. We first treat the Hermitian defect, then briefly discuss the non‑Hermitian variant.

The Burkov–Balents Hamiltonian with a single off‑diagonal defect in layer~$I$ is
\begin{gather}
\mathcal{H}_{sd}=\mathcal{H}_{0}+\delta\mathcal{H},
\end{gather}
where
\begin{gather}
\mathcal{H}_{0}=\sum_{\mathbf{k}_{\perp},i,j}
c_{\mathbf{k}_{\perp}i}^{\dagger}\,
T_{ij}\!\left(\mathbf{k}_{\perp}\right)\,
c_{\mathbf{k}_{\perp}j},\\
\delta\mathcal{H}=\sum_{\mathbf{k}_{\perp}}
c_{\mathbf{k}_{\perp}I}^{\dagger}\,\delta\hat{h}\,
c_{\mathbf{k}_{\perp}I},
\end{gather}
with $\delta\hat{h}=\tau^{x}\delta\Delta_{S}$ and
$\delta\Delta_{S}=\tilde{\Delta}_{S}-\Delta_{S}$.  
The label \textit{sd} stands for “single defect”.

\begin{figure}[b!]
\centering
\includegraphics[width=\columnwidth]{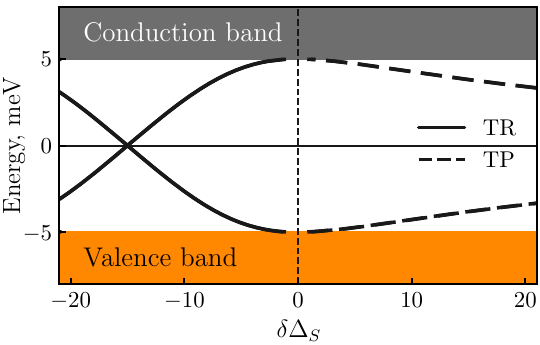}
\caption{Energy of the defect–induced bound state as a function of the tunneling variation $\delta\Delta_{S}$. The solid line (TR) corresponds to the trivial phase ($\Delta_{S}>\Delta_{D}$), while the dashed line (TP) corresponds to the topological phase ($\Delta_{S}<\Delta_{D}$). The grey and orange shaded regions mark the conduction and valence bands, respectively; the horizontal line at $E=0$ is the mid–gap energy, and the vertical dashed line at $\delta\Delta_{S}=0$ indicates the clean limit. In the trivial phase the bound state emerges in the gap only for $\delta\Delta_{S}<0$ and crosses mid–gap at $\delta\Delta_{S}=-\Delta_{S}$; in the topological phase it appears only for $\delta\Delta_{S}>0$ and never reaches zero energy. The curves are obtained from the pole condition, Eq.~\eqref{poles condition}. TP stands for ``Topological'', TR for ``Trivial''.}
\label{bandgap states}
\end{figure}

Expanding the Green function $G_{ij}\bigl(\mathbf{k}_{\perp}\bigr)$ in powers of the perturbation $\delta\hat{h}$ yields, by analogy with the standard $t$‑matrix procedure, the Dyson equation
\begin{equation}
G_{ij}^{sd}\!\left(\mathbf{k}_{\perp}\right)
=G_{ij}^{0}\!\left(\mathbf{k}_{\perp}\right)
+G_{iI}^{0}\!\left(\mathbf{k}_{\perp}\right)
\delta\hat{h}\,
G_{Ij}^{sd}\!\left(\mathbf{k}_{\perp}\right).
\label{Dyson_single-defect}
\end{equation}
Rewriting Eq.~\eqref{Dyson_single-defect} gives
\begin{multline}
G_{ij}^{sd}\!\left(\mathbf{k}_{\perp}\right)
=G_{ij}^{0}\!\left(\mathbf{k}_{\perp}\right)\\
+G_{iI}^{0}\!\left(\mathbf{k}_{\perp}\right)
\delta\hat{h}\,
\bigl[\,1-G^{0}\!\left(\mathbf{k}_{\perp}\right)\delta\hat{h}\,\bigr]^{-1}
G_{Ij}^{0}\!\left(\mathbf{k}_{\perp}\right).
\label{sd Green function}
\end{multline}
Poles of $G_{ij}^{sd}$ correspond to the eigenvalues of~$\mathcal{H}_{sd}$.  
From Eq.~\eqref{sd Green function} the pole condition is
\begin{equation}
\det\!\bigl[1-G^{0}\!\left(\mathbf{k}_{\perp}\right)\delta\hat{h}\bigr]=0.
\label{Green function pole}
\end{equation}
Inside the bands $G^{0}$ has an imaginary part, so Eq.~\eqref{Green function pole} splits into two incompatible real equations and no new state appears.  
Within the gap $G^{0}$ is real, the matrix $1-\delta\hat{h}G^{0}$ can be diagonalised, and Eq.~\eqref{Green function pole} reduces to
\begin{equation}
\prod_{i}\lambda_{i}=0,
\end{equation}
where $\lambda_{i}$ are the eigenvalues of $1-\delta\hat{h}G^{0}$.  
We focus on the region $\varepsilon_{\perp}^{2}<(\Delta_{S}-\Delta_{D})^{2}$.  
Using the local Green function from Appendix~B, Eq.~\eqref{Green_ll}, we write
\begin{gather}
G_{ll}^{0}=G_{0}+G_{x}\tau^{x}+G_{z}k_{y}\,\tau^{z}\!\otimes\!\sigma_{x}
          -G_{z}k_{x}\,\tau^{z}\!\otimes\!\sigma_{y}.
\label{G_ll_matrix}
\end{gather}
The poles are then given by
\begin{gather}
1-\delta\Delta_{S}G_{x}\pm
\delta\Delta_{S}\bigl|\,\sqrt{G_{0}^{2}-G_{z}^{2}k_{\perp}^{2}}\,\bigr|=0.
\label{poles condition}
\end{gather}
There are two solutions to this equation: $\varepsilon_{1,2}^0$. Because \eqref{poles condition} depends on $\mathbf{k}_{\perp}$ only through $\varepsilon_{\perp}$, we consider the time‑reversal point $\mathbf{k}_{\perp}=0$ for simplicity.  
No solutions exist inside the allowed bands.  
The equation holds at $\varepsilon=-\Delta_{S}\pm\Delta_{D}$ and $\varepsilon=\Delta_{S}\pm\Delta_{D}$--the band edges--for any $\delta\Delta_{S}$.  
Within the gap two solutions symmetric about zero energy appear.  
Figure~\ref{bandgap states} shows how their positions vary with $\delta\Delta_{S}$.  
New in‑gap states arise in both trivial and topological phases, yet with a clear difference:  
they occur for $\delta\Delta_{S}<0$ in the trivial phase and for $\delta\Delta_{S}>0$ in the topological phase.  
Moreover, in the trivial phase the bound state crosses mid‑gap at $\delta\Delta_{S}=-\Delta_{S}$, whereas in the topological phase it never reache    s zero energy.  
Physically, $\delta\Delta_{S}=-\Delta_{S}$ sets the tunneling in the defect layer to zero, effectively splitting the stack into two parts; the defect’s edge modes then become boundary modes of each half and sit at zero energy in the trivial phase.  
In the topological phase ($\Delta_{S}<\Delta_{D}$), Eq.~\eqref{poles condition} implies $\delta\Delta_{S}/(\Delta_S+\delta\Delta_{S})=1$ only as $\delta\Delta_{S}\to\infty$, so the extra state never crosses zero.

Next we evaluate the degeneracy $d_{f}$ of the in‑gap state $\varepsilon_{2}^{0}$.  
It is given by the residue of the Green function:
\begin{equation}
d_{f}=\mathrm{Tr}\!\bigl[\mathrm{Res}\,G\bigl(\varepsilon_{2}^{0}\bigr)\bigr].
\end{equation}
At $\varepsilon=\varepsilon_{2}^{0}$ the residue is
\begin{equation}
\mathrm{Res}\,G\bigl(\varepsilon_{2}^{0}\bigr)
=G_{iI}^{0}\!\bigl(\varepsilon_{2}^{0}\bigr)\,
\delta h\,\,
\frac{\mathrm{adj}\!\bigl[1-\delta h\,G^{0}\!\bigl(\varepsilon_{2}^{0}\bigr)\bigr]}
     {\det'\!\bigl[1-\delta h\,G^{0}\!\bigl(\varepsilon_{2}^{0}\bigr)\bigr]}\,
G_{Ij}^{0}\!\bigl(\varepsilon_{2}^{0}\bigr),
\label{Residue}
\end{equation}
so that
\begin{multline}
d_{f}
=\sum_{n}
\langle n|G_{\alpha\beta}^{0}|I\rangle
\,\delta h^{\beta\gamma}M_{\gamma\lambda}\,
\langle I|G_{\lambda\alpha}^{0}|n\rangle
\\
=\langle I|G_{\lambda\alpha}^{0}G_{\alpha\beta}^{0}|I\rangle
\,\delta h^{\beta\gamma}M_{\gamma\lambda}
=\langle I|\bigl(G^{0}\bigr)_{\lambda\beta}^{2}|I\rangle
\,\delta h^{\beta\gamma}M_{\gamma\lambda},
\end{multline}
where
$
M_{\gamma\lambda}
=\mathrm{adj}[1-\delta h\,G^{0}(\varepsilon_{2}^{0})]_{\gamma\lambda}
/\det'[1-\delta h\,G^{0}(\varepsilon_{2}^{0})]
$
and repeated indices $\alpha,\beta,\gamma,\lambda$ are summed.  
One finds $\delta h^{\beta\gamma}M_{\gamma\lambda}
=-(\partial_{\varepsilon}G_{\beta\lambda}^{0})^{-1}_{\varepsilon=\varepsilon_{2}^{0}}$,  
so $d_{f}=1$ for a given spin projection, since $(G^{0})^{2}=-\partial_{\varepsilon}G^{0}$.  
Equation~\eqref{Residue} also shows that the new state is localized near layer~$I$:  
using $\langle n|b\rangle\langle b|m\rangle=\mathrm{Res}\,G_{nm}^{sd}(\varepsilon_{2}^{0})$ gives  
$\langle n|b\rangle\propto G_{nI}^{0}\sim e^{-aR_{nI}}$ inside the gap.

\begin{figure}[t!]
\centering
\includegraphics[width=\columnwidth]{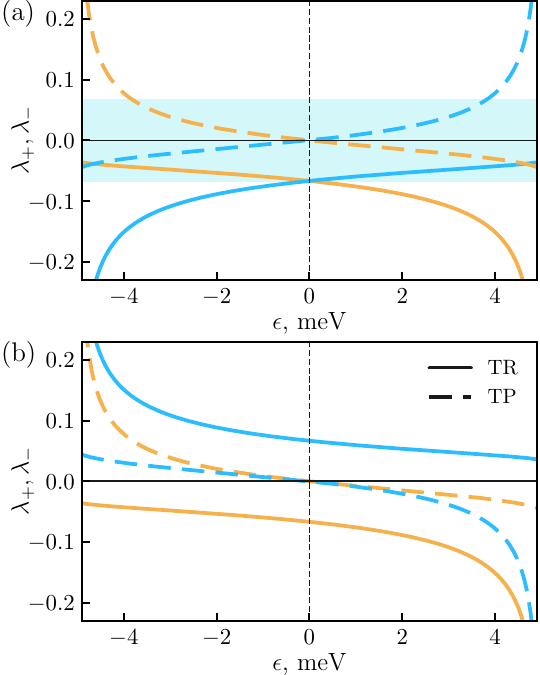}
\caption{Energy dependence of the two eigenvalues $\lambda_{\pm}$ defined in Eq.~\eqref{Green function eigenvalues} for the \emph{local} Green operator. Panel~(a) shows the case of off–diagonal disorder (a single tunneling defect), panel~(b) the case of diagonal disorder. Solid curves (TR) correspond to the trivial phase $(\Delta_{S}>\Delta_{D})$, dashed curves (TP) to the topological phase $(\Delta_{S}<\Delta_{D})$. The vertical dashed line marks $\varepsilon=0$. In panel~(a) both eigenvalues have the same sign inside the gap in the trivial phase, whereas in the topological phase they approach zero at mid–gap; as a consequence, the pole condition $1/\delta\Delta_{S}=\lambda_{\pm}(\varepsilon)$ (illustrated by the shaded horizontal strip) can be satisfied only for a definite sign of the tunneling variation, and the bound state never crosses zero energy in the topological regime. In contrast, in panel~(b) (diagonal disorder) one finds $\lambda_{+}\lambda_{-}<0$ throughout the gap, so a crossing with $1/\delta V$ ($V$ is the on–site perturbation) always occurs for one of the two eigenvalues, recovering the familiar behavior for diagonal impurities. TP stands for ``Topological'', TR for ``Trivial''.}
\label{Green eigenvalues}
\end{figure}

We now examine the topological character of bound states created by a single off‑diagonal defect.  
For diagonal disorder of codimension one (point impurity in 1D, line in 2D, etc.), Ref.~\cite{Slager} showed that a bound state always appears in the topological phase, whereas in the trivial phase it depends on details.  
The key is that eigenvalues of the local Green operator cross zero only in the topological phase.  
Here the disorder is again of codimension one (a defective layer) but off‑diagonal in nature, so the conclusions of Ref.~\cite{Slager} do not carry over directly.  
We therefore compute the eigenvalues of $\tau^{x}G^{0}_{ll}$,
\begin{gather}
\lambda_{\pm}=G_{x}(\varepsilon)\pm\bigl|G_{0}(\varepsilon)\bigr|.
\label{Green function eigenvalues}
\end{gather}
Their behavior is plotted in Fig.~\ref{Green eigenvalues}.  
For $\varepsilon^{2}<(\Delta_{S}-\Delta_{D})^{2}$ with $\Delta_{S}>\Delta_{D}$ one has $|G_{x}|>|G_{0}|$, so $\lambda_{\pm}>0$--the chief difference from diagonal disorder, where $\lambda_{+}\lambda_{-}<0$.  
Hence $\lambda_{\pm}=1/\delta\Delta_{S}$ can be satisfied only for one sign of $\delta\Delta_{S}$: in the trivial phase ($\Delta_{S}>\Delta_{D}$) an in‑gap state arises only for a specific sign of the perturbation.  
At $\varepsilon=0$, $G_{0}=0$ and $\lambda_{+}=\lambda_{-}=-1/\Delta_{S}$ in the trivial phase, so the bound state crosses zero at $\delta\Delta_{S}=-\Delta_{S}$.  
In the topological phase $\lambda_{+}=\lambda_{-}=0$ at $\varepsilon=0$, and the crossing never occurs.

If the perturbation also shifts $\Delta_{D}$ between layers $I$ and $I-1$, the defect Hamiltonian becomes
\begin{multline}
\delta\mathcal{H}=\sum_{\mathbf{k}_{\perp}}
\Bigl[
\delta\Delta_{S}\,c_{\mathbf{k}_{\perp}I}^{\dagger}\tau^{x}c_{\mathbf{k}_{\perp}I}\\
+\frac{\delta\Delta_{D}}{2}
\bigl(c_{\mathbf{k}_{\perp}I-1}^{\dagger}\tau^{+}c_{\mathbf{k}_{\perp}I}
+c_{\mathbf{k}_{\perp}I}^{\dagger}\tau^{-}c_{\mathbf{k}_{\perp}I-1}\bigr)
\Bigr].
\end{multline}
Analytical treatment then becomes more involved because interlayer Green functions enter via the off‑diagonal matrix elements.

\subsection{Non-Hermitian}

\begin{figure}
\centering
\includegraphics[width=1\columnwidth]{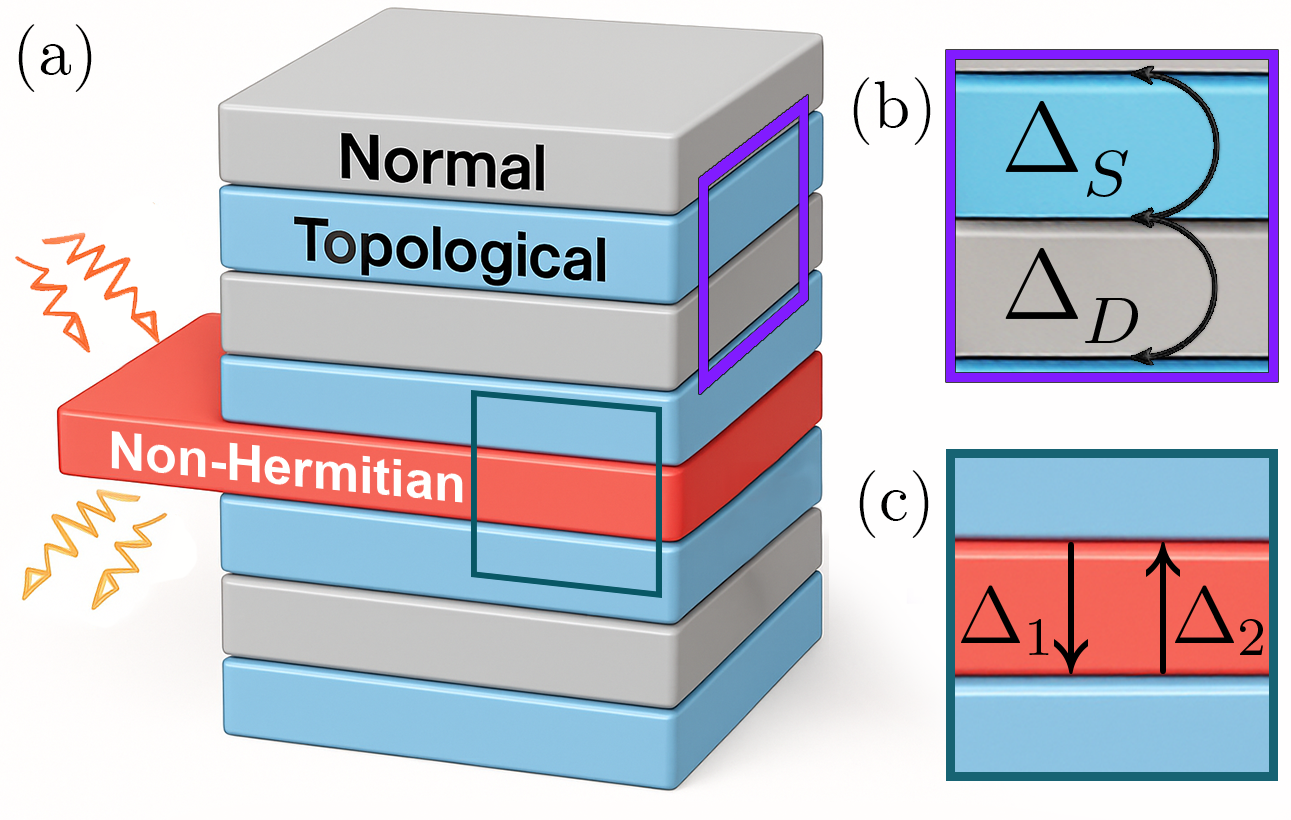}
\caption{(a) Van der Waals multilayer composed of alternating topological–insulator (blue) and trivial (grey) films. A single \emph{non‑Hermitian} TI layer (red) is coupled to an external environment (wiggly arrows), which endows it with a finite lifetime and renders the effective Hamiltonian non‑Hermitian. (b) Zoom on a \emph{Hermitian} (clean) interface: the two intra–layer tunneling amplitudes that enter the Burkov–Balents model are indicated, $\Delta_{S}$ (through a TI film) and $\Delta_{D}$ (across a NI spacer). (c) Zoom on the defective layer: besides the imaginary on–site potential $i\gamma$ (not shown explicitly), forward and backward hoppings on that layer become unequal, $\Delta_{1}\neq\Delta_{2}$, producing asymmetric interlayer coupling. These two ingredients realise the perturbation of Eq.~\eqref{nonHermitian defect}, allowing us to study both ``homogeneous'' ($\tilde{h}=\tau^{0}i\gamma$) and ``inhomogeneous'' ($\tilde{h}=\tau^{z}i\gamma$) non‑Hermiticity while keeping the rest of the stack Hermitian.}
\label{non-Hermitian layer}
\end{figure}

We now examine how non‑Hermitian effects modify the single‑defect in‑gap state.  
The set‑up is sketched in Fig.~\ref{non-Hermitian layer}: the defect layer itself is made non‑Hermitian.  
Two sources of non‑Hermiticity are included:  
(i) an imaginary on‑site potential, representing the finite lifetime of electrons in that layer, and  
(ii) an asymmetry between forward and backward hopping on the defect layer.  
Such terms may arise from coupling the layer to an external environment, as indicated in the figure.

The resulting generalised Burkov–Balents Hamiltonian is
\begin{gather}
\mathcal{H}_{\mathrm{n\text{-}H}}=\mathcal{H}_{0}+\delta\mathcal{H},
\end{gather}
with
\begin{gather}
\delta\mathcal{H}
=c_{\mathbf{k}_{\perp}I}^{\dagger}
\bigl(\tilde{h}
      +\tau^{+}\Delta_{1}
      +\tau^{-}\Delta_{2}
      -\tau^{x}\Delta_{S}\bigr)
c_{\mathbf{k}_{\perp}I}.
\label{nonHermitian defect}
\end{gather}
In general this perturbation breaks $\mathcal{R}$, $\mathcal{T}$, and $\mathcal{RT}$ symmetries.  
When $\Delta_{1}=\Delta_{2}\neq\Delta_{S}$ and $\tilde{h}=0$ the Hamiltonian is Hermitian and reduces to the single off‑diagonal defect considered earlier;  
the defect disappears for $\Delta_{1}=\Delta_{2}=\Delta_{S}$.  

We consider two choices for the imaginary potential $\tilde{h}$:
\begin{gather}
\tilde{h}=
\begin{cases}
\tau^{0}\,i\gamma,\\[4pt]
\tau^{z}\,i\gamma,
\end{cases}
\label{Two forms of nonHermit}
\end{gather}
corresponding to (i) equal loss (or gain) on the top and bottom surfaces of the TI layer, and  
(ii) different loss/gain on the two surfaces.  
We refer to case~(i) as \emph{homogeneous} non‑Hermiticity and case~(ii) as \emph{inhomogeneous}.  
Both scenarios are analyzed below.

The perturbation in Eq.~\eqref{nonHermitian defect} breaks the $\mathcal{R}$, $\mathcal{T}$, and $\mathcal{RT}$ symmetries in the case of \emph{homogeneous} non‑Hermiticity ($\tilde{h}=\tau^{0} i\gamma$).  
For \emph{inhomogeneous} non‑Hermiticity ($\tilde{h}= \tau^{z} i\gamma$) the $\mathcal{R}$ and $\mathcal{T}$ symmetries are each broken separately, yet the combined $\mathcal{RT}$ symmetry remains intact: $(\mathcal{RT})^{-1}\tilde{h}\,\mathcal{RT}= \tilde{h}$, the analogue of $\mathcal{PT}$ symmetry.

Note that Hamiltonian~\eqref{nonHermitian defect} retains chiral symmetry in \emph{both} cases~\eqref{Two forms of nonHermit}.  
The chirality operator may be written as the product of the time‑reversal operator $\mathcal{T}=i\tau^{0}\!\otimes\!\sigma_{y}K$ and $\mathcal{K}=\tau^{z}\!\otimes\!\sigma_{x}K$.  
For the full Hamiltonian $\mathcal{H}_{0}+\delta\mathcal{H}$ this becomes an electron–hole symmetry, implying a spectrum symmetric about zero energy.

For later use we recast Eq.~\eqref{nonHermitian defect} in a compact form,
\begin{gather}
\delta\mathcal{H}
=c_{\mathbf{k}_{\perp}I}^{\dagger}
\bigl(\tilde{h}+\tau^{x}\delta\Delta_{S}+\tau^{-}\delta\Delta_{N}\bigr)
c_{\mathbf{k}_{\perp}I},
\label{nonHermitian defect+}
\end{gather}
where $\delta\Delta_{N}\equiv\Delta_{2}-\Delta_{1}$.  
With non‑Hermitian terms present, the pole condition reads
\begin{gather}
\mathrm{Re}\!\bigl[\det\bigl(1-\widehat{G}^{0}_{ll}\,\delta\hat{h}\bigr)\bigr]=0 .
\label{polus nonHermit}
\end{gather}
Using the unperturbed local Green function~\eqref{G_ll_matrix}, Eq.~\eqref{polus nonHermit} can be solved analytically for both choices in~\eqref{Two forms of nonHermit}.  For $k_{z}=0$ we obtain:

\noindent \emph{(i) Homogeneous non‑Hermiticity} ($\tilde{h}=\tau^{0} i\gamma$):
\begin{multline}
0=\pm 2\gamma\,G_{0}
-G_{0}^{2}\!\left(\gamma^{2}+\delta\Delta_{S}(\delta\Delta_{N}+\delta\Delta_{S})\right)
\\
+\Bigl[1-G_{x}\bigl(\tfrac{\delta\Delta_{N}}{2}+\delta\Delta_{S}\bigr)\Bigr]^{2}
+G_{x}^{2}\!\left(\gamma^{2}-\tfrac{\delta\Delta_{N}^{2}}{4}\right);
\label{homogeneous nonhermicity}
\end{multline}

\begin{figure}[t!]
\centering
\includegraphics[width=1\columnwidth]{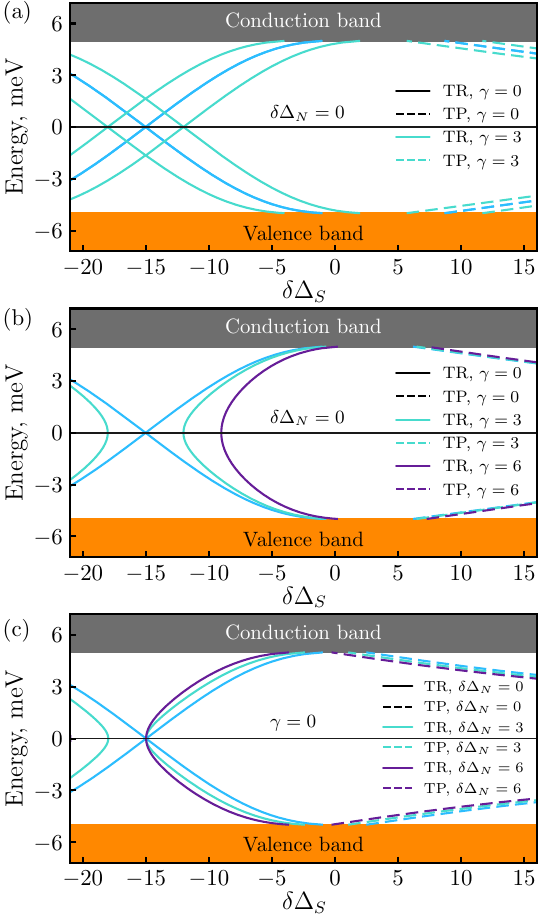}
\caption{Bound–state energies generated by a single tunneling defect as a function of $\delta\Delta_{S}$ for several types and strengths of non-Hermiticity. Solid (dashed) curves correspond to the trivial (TR) and topological (TP) phases, respectively; all lines are obtained from Eqs.~\eqref{homogeneous nonhermicity} and \eqref{inhomogeneous nonhermicity}. (a) Homogeneous loss/gain, $\tilde{h}=\tau^{0} i\gamma$ with $\delta\Delta_{N}=0$: increasing $\gamma$ splits and shifts the in‑gap levels but the characteristic rule persists--the level crosses zero only in the trivial phase. (b) Inhomogeneous loss/gain, $\tilde{h}=\tau^{z} i\gamma$: the remaining $\mathcal{RT}$ symmetry preserves a twofold degeneracy; larger $\gamma$ bends the branches without enabling a zero‑energy crossing in the TP regime. (c) Purely asymmetric hopping, $\gamma=0$ and $\delta\Delta_{N}\neq0$: nonreciprocity alone produces a similar splitting and displacement of the bound state. Grey (orange) shading marks the conduction (valence) band continua.}
\label{non-Hermitian bandgap states}
\end{figure}

\noindent  \emph{(ii) Inhomogeneous non‑Hermiticity} ($\tilde{h}=\tau^{z} i\gamma$):
\begin{multline}
\bigl(G_{0}^{2}-G_{x}^{2}\bigr)
\bigl(\gamma^{2}-\delta\Delta_{S}(\delta\Delta_{N}+\delta\Delta_{S})\bigr)
\\
-G_{x}\!\bigl(\delta\Delta_{N}+2\delta\Delta_{S}\bigr)+1=0 .
\label{inhomogeneous nonhermicity}
\end{multline}
The resulting solutions, plotted as functions of $\delta\Delta_S$, are shown in Fig.~\ref{non-Hermitian bandgap states}. for different types of non-Hermiticity. Introducing a non‑Hermitian component lifts the degeneracy of the bound‑state levels with respect to both real spin and pseudospin (upper vs. lower TI surface), reflecting the broken $\mathcal{R}$ and $\mathcal{T}$ symmetries.  
For inhomogeneous non‑Hermiticity, however, the $\mathcal{RT}$ symmetry preserves a two‑fold degeneracy.

A key result is that the in‑gap bound state in the \emph{topological} phase is essentially insensitive to the non‑Hermitian perturbation.

\section{Off‑diagonal disorder: density of states and localization}

Having analyzed the single‑defect problem, we now turn to the case in which off‑diagonal disorder is distributed throughout the sample with a finite concentration.

\subsection{Density of states, chemical potential, and Hall resistivity}

We first investigate how off‑diagonal disorder modifies the density of states (DOS) and the chemical potential, using the two approaches developed in Sec.~3.  
Numerical results are presented below.  The DOS can be written as
\begin{gather}
\rho(\epsilon)
=-\frac{1}{\pi}\,\mathrm{Im}
\sum_{\mathbf{k}_{\perp},k_{z}}
\mathrm{tr}\Bigl\langle
G_{k_{z}}\!\bigl(\mathbf{k}_{\perp},\epsilon+i0\bigr)
\Bigr\rangle ,
\label{density of states}
\end{gather}
so that Eqs.~\eqref{locator}, \eqref{interactor}, \eqref{self_green_func}, and~\eqref{GF t-matrix} allow us to compute $\rho(\epsilon)$.  
The renormalized chemical potential $\overline{\mu}$ is determined from
\begin{gather}
\int_{0}^{\overline{\mu}}\rho(\epsilon)\,d\epsilon
=\int_{0}^{\mu}\rho_{0}(\epsilon)\,d\epsilon,
\label{eq: chem potential}
\end{gather}
where $\rho_{0}$ is the DOS of the clean system.

All calculations are performed with both methods discussed in Sec.~3, namely the expansion in localized states and the expansion in Bloch states.  
For each method we also present results obtained in a zeroth‑order, non‑self‑consistent approximation for reference.

In the locator approximation we set
\begin{gather}
\overline{t}=0,\\
\sigma=\sigma_{0}=S_{m},\\
\bigl\langle G^{1}_{k_{z}}\bigl(\mathbf{k},\epsilon\bigr)\bigr\rangle
=\Bigl(\langle\sigma_{0}\rangle^{-1}-T_{k_{z}}\Bigr)^{-1}.
\label{locator_approx}
\end{gather}

The analogous zeroth‑order Bloch‑state treatment reads
\begin{gather}
\overline{\upsilon}
=\overline{\upsilon}_{0}
=\overline{t}_{0}\bigl(1+G_{l}^{0}\overline{t}_{0}\bigr)^{-1},\\
\overline{t}_{0}
=\Bigl\langle\bigl(1-V_{l}G_{ll}^{0}\bigr)^{-1}V_{l}\Bigr\rangle ,\\
\bigl\langle G^{1}_{k_{z}}\bigl(\mathbf{k},\epsilon\bigr)\bigr\rangle
=\Bigl(\epsilon-\overline{\upsilon}_{0}-T_{k_{z}}\bigl(\mathbf{k}\bigr)\Bigr)^{-1}.
\label{band state_approx}
\end{gather}

\begin{figure}[b!]
    \centering
    \includegraphics[width=\columnwidth]{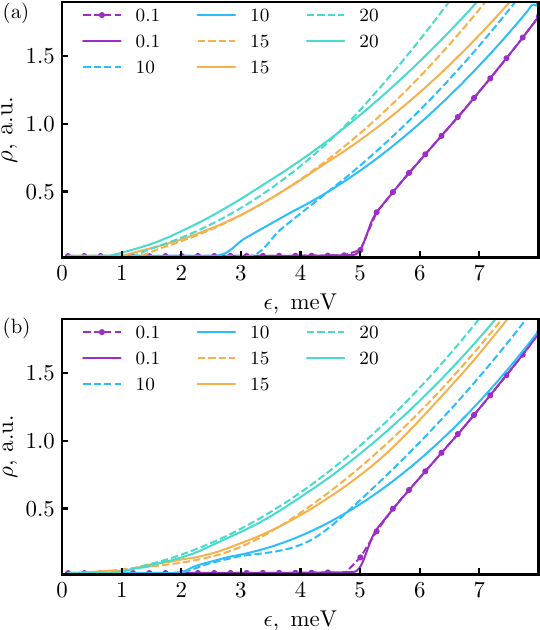}
    \caption{Density of states $\rho(\varepsilon)$ in the \textit{trivial–insulator} regime for several amplitudes of off–diagonal disorder $\eta_{0}$ (numbers in the legend, in meV). Solid lines are obtained within the self–consistent locator scheme, dashed lines from the Bloch–state ($t$–matrix+RPA) expansion. (a) Fully self–consistent implementation of the two methods; (b) their non–self–consistent (zeroth–order) versions, Eqs.~\eqref{locator_approx} and \eqref{band state_approx}. In both treatments increasing $\eta_{0}$ produces a Lifshitz–type tail inside the gap and pushes the band edge towards lower energies, eventually narrowing the gap substantially; the approximate curves slightly overestimate this narrowing but remain close to the self–consistent result.}
    \label{dos_triv}
\end{figure}

\begin{figure}
    \centering
    \includegraphics[width=\columnwidth]{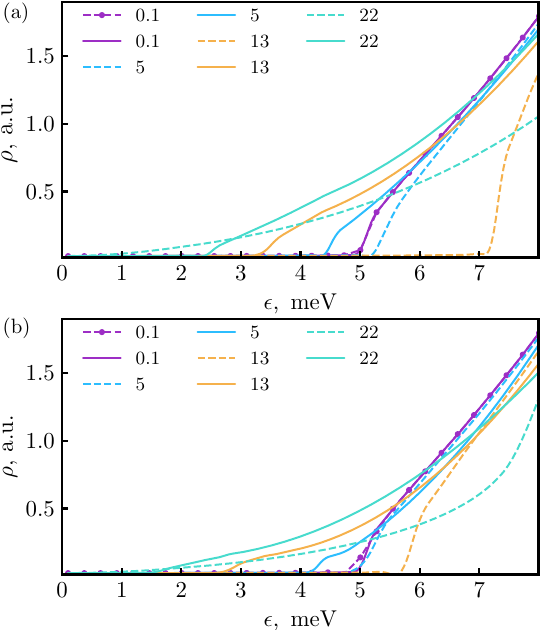}
        \caption{Density of states $\rho(\varepsilon)$ of the multilayer topological insulator in the \textit{topological} phase for several amplitudes of off–diagonal disorder $\eta_{0}$ (values in meV are indicated in the legend). Solid lines: self–consistent locator scheme; dashed lines: Bloch–state ($t$–matrix+RPA) expansion. Panel (a) shows the fully self–consistent implementation of the two methods, while panel (b) displays their non–self–consistent (zeroth–order) versions, Eqs.~\eqref{locator_approx} and \eqref{band state_approx}. For weak disorder the two approaches differ qualitatively: the locator method predicts a slight \emph{reduction} of the gap, whereas the Bloch–state expansion yields a modest \emph{increase}. For stronger $\eta_{0}$ both schemes generate pronounced in–gap (Lifshitz–tail) states and eventually drive a collapse of the gap.}
    \label{dos_topo}
\end{figure}

The numerical results are collected in Figs.~\ref{dos_triv} and \ref{dos_topo}.  
In the trivial phase both approaches give almost identical curves: disorder narrows the gap by creating a fluctuation tail, and the zeroth‑order approximation (Fig.~\ref{dos_triv}b) slightly overestimates this narrowing but remains close to the self‑consistent result (Fig.~\ref{dos_triv}a).  
For sufficiently strong disorder the gap eventually collapses.  

In the topological phase (Fig.~\ref{dos_topo}) strong disorder likewise drives a gap collapse, consistent across both self-consistent and approximate schemes.  
At weak disorder, however, the two methods diverge: the locator scheme predicts a \emph{reduction} of the gap width, whereas the Bloch‑state expansion yields a modest \emph{increase}.  
The origin of this discrepancy is not yet understood and will be explored elsewhere.  

\begin{figure}
    \centering
    \includegraphics[width=\columnwidth]{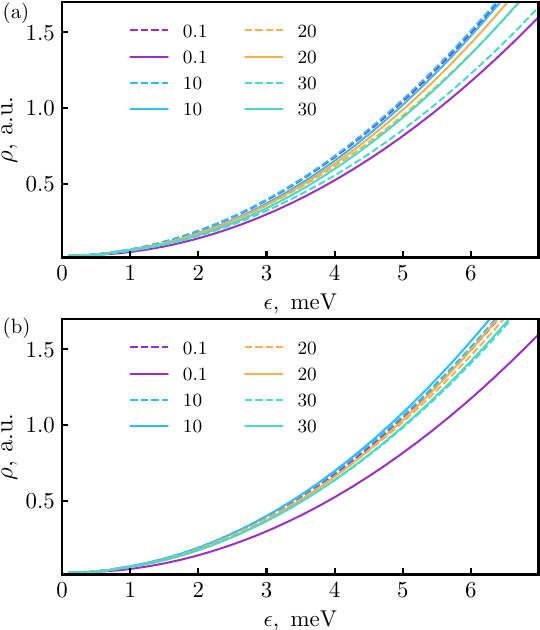}
    \caption{Density of states $\rho(\varepsilon)$ of the multilayer system in the \textit{Weyl semimetal} phase for several amplitudes of off–diagonal disorder $\eta_{0}$ (numbers in meV are given in the legend). Solid lines correspond to the self–consistent locator scheme, dashed lines to the Bloch–state ($t$–matrix+RPA) expansion. Panel (a) shows the fully self–consistent implementation, panel (b) the zeroth–order (non–self–consistent) versions. In contrast to the insulating phases, all curves remain gapless and nearly coincide between the two methods, indicating that the Weyl phase is robust: disorder merely renormalizes the low–energy slope of $\rho(\varepsilon)$ without opening a gap.}
    \label{dos_weyl}
\end{figure}

\begin{figure}
    \centering
    \includegraphics[width=\columnwidth]{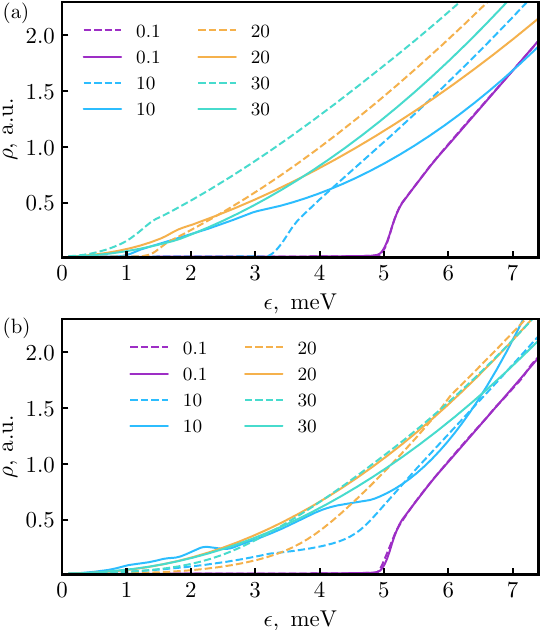}
    \caption{Density of states $\rho(\varepsilon)$ in the \textit{anomalous quantum Hall} (AQHE) phase for several amplitudes of off–diagonal disorder $\eta_{0}$ (values in meV are indicated in the legend). Solid curves – self–consistent locator scheme; dashed curves – Bloch–state ($t$–matrix+RPA) expansion. (a) Fully self–consistent implementation; (b) the corresponding zeroth–order (non–self–consistent) approximations. Increasing $\eta_{0}$ produces pronounced fluctuation tails that leak into the gap and eventually close it, shrinking the Hall plateau. In this phase the two methods agree quantitatively over the whole range of disorder.}
    \label{dos_AQHE}
\end{figure}

Overall, the calculations show that substantial off‑diagonal disorder generates in‑gap bulk states and can even close the gap entirely.  
Such states may influence edge transport and should be regarded as an additional mechanism by which measured edge conductance deviates from the ideal quantized value.  
Throughout Figs.~\ref{dos_triv}–\ref{dos_AQHE}, panel~(a) displays the fully self‑consistent result, while panel~(b) corresponds to the approximate treatments defined in Eqs.~\eqref{locator_approx} and \eqref{band state_approx}.

Adding a Zeeman term, $S_{j}^{-1}\!\to
S_{j}^{-1}-\tau^{0}\sigma_{z}\Delta_{Z}$, allows us to test how
off‑diagonal disorder affects the stability of the Weyl and AQHE phases.
Figure~\ref{dos_weyl} displays the DOS in the Weyl phase, while
Fig.~\ref{dos_AQHE} shows the DOS in the AQHE phase, both for various
values of $\eta_{0}$.

\begin{figure}
    \centering
    \includegraphics[width=\columnwidth]{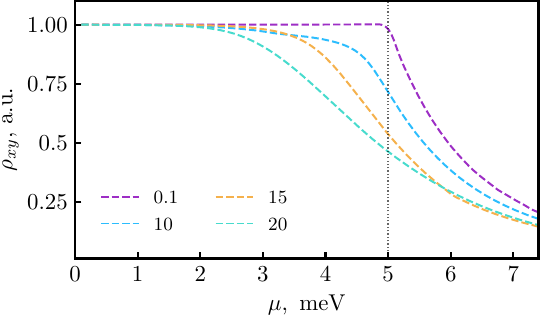}
    \caption{Hall resistivity $\rho_{xy}$ (arbitrary units) in the AQHE phase versus the chemical potential $\mu$ for several amplitudes of off–diagonal disorder $\eta_{0}$ (values in meV are indicated in the legend). Curves are evaluated from Eq.~\eqref{Hall resistivity} using the disorder–broadened DOS of Fig.~\ref{dos_AQHE}. The quantized plateau at $\rho_{xy}=1$ survives for weak disorder and $\mu$ inside the clean gap (vertical dotted line), but is progressively eroded and narrowed as $\eta_{0}$ increases: fluctuation–induced in‑gap bulk states enhance $\sigma_{b}$, driving $\rho_{xy}$ away from its universal value and pushing the crossover to smaller $\mu$.}
    \label{resistance_Hall}
\end{figure}

In the AQHE phase the two self‑consistent schemes agree quantitatively, and even the zeroth‑order approximation performs well.  
Off‑diagonal disorder again narrows the effective band gap and produces additional bulk states inside it.  
Both effects alter the Hall response, which can be illustrated with a simple Drude picture.  
The Hall resistivity is written as
\begin{equation}
\rho_{H}=\frac{\sigma_{e}}{\sigma_{e}^{2}+\sigma_{b}^{2}},
\label{Hall resistivity}
\end{equation}
where $\sigma_{e}=e^{2}/\hbar$ is the quantized conductance of the chiral edge modes and $\sigma_{b}$ is the bulk conductance.  
When the Fermi level lies inside the gap in an ideal AQHE system, $\sigma_{b}=0$.  
Figure~\ref{resistance_Hall} plots $\rho_{H}$ versus the Fermi energy $\mu$ for several disorder strengths $\eta_{0}$.  
As the figure shows, off‑diagonal disorder partially erodes the Hall plateau, whose width decreases with increasing $\eta_{0}$.

\begin{figure}
    \centering
    \includegraphics[width=1\linewidth]{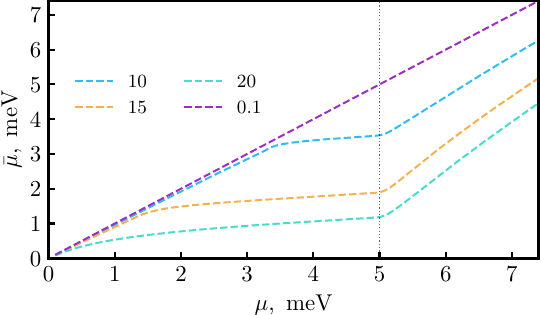}
    \caption{Disorder–induced renormalization of the chemical potential: $\overline{\mu}$ (defined by Eq.~\eqref{eq: chem potential}) plotted versus the bare $\mu$ for several amplitudes of off–diagonal disorder $\eta_{0}$ (values in meV are shown in the legend).    In the clean or weak–disorder limit ($\eta_{0}=0.1$) $\overline{\mu}\!\approx\!\mu$.    As $\eta_{0}$ increases, spectral weight is transferred into the gap and
    $\overline{\mu}$ is pushed down, producing pronounced kinks near the clean gap edge
    (vertical dotted line) and extended plateaus where the DOS is finite already below the band edge.}
    \label{chem potential}
\end{figure}

Figure~\ref{chem potential} displays the renormalized chemical potential
$\overline{\mu}(\mu)$ in the topological‑insulator phase, evaluated from
Eq.~\eqref{eq: chem potential}.  
The plot shows that the chemical potential of the disordered system
shifts inside the fluctuation region where bulk in‑gap states emerge.

Work~\cite{Ovchinnikov1977} (see also~\cite{gredeskul1978state})
predicted that fluctuations of the gap width--i.e.\ off‑diagonal disorder--produce
a singularity in the DOS at mid‑gap, an effect first analyzed by
Dyson~\cite{Dyson}.  In the present three‑dimensional system the DOS at
zero energy remains regular.  As argued in
Ref.~\cite{Alisultanov_multilayer}, this likely stems from both the
dimensionality and the perturbative nature of our methods, a conclusion
we maintain here.  Dyson’s model shows that such a singularity persists
for a uniform distribution, whereas our calculations do not reveal it.
In the next subsection we therefore examine off‑diagonal disorder at
zero energy in greater detail, focusing on the localization length of
the TI edge modes and on the possibility of disorder‑induced
localization in the gapless (Dirac) phase.

\subsection{The simplest analysis of localization}

A principal consequence of disorder is the localization of electronic states.  
This phenomenon was first predicted by P.~W.~Anderson in his seminal paper on diagonal disorder~\cite{anderson1958absence}.  
Several later works have reported localization induced by off‑diagonal disorder as well (see, e.g.,~\cite{Raghavan,Harris,Timothy_Ziman}).  
Here we address this issue for a multilayer topological insulator and ask whether off‑diagonal disorder can localize states in the gapless Dirac or Weyl phase and, conversely, delocalize them in the gapped topological‑insulator phase.  
Our analysis focuses mainly on states at zero energy.

The central result of Anderson’s original study~\cite{anderson1958absence} is a theorem stating that, above a certain disorder strength, \emph{all} states in the system become localized.  
Mathematically, part of the proof reduces to showing the convergence of a series for the Green function (see the discussion in~\cite{anderson1958absence} and in~\cite{economou1972existence}).  
A full, rigorous proof is quite involved and lies beyond the scope of this work.  
Instead, we adopt a simpler approach proposed by Ziman~\cite{J.M.Ziman}.  
Although approximate, Ziman’s method provides a convenient way to assess the key localization criterion and is well suited for the qualitative analysis required here.

We first address the localization length of edge modes in the
topological‑insulator phase in the presence of off‑diagonal disorder.
The qualitative argument in Sec.~2 showed that such disorder drives the
localization length to infinity, i.e.\ delocalizes the modes.  Here we
examine the point in more detail using the Green‑function formalism.  The
expansion derived in Appendix~A, Eq.~\eqref{Green perturb series}, reads
\begin{gather}
G_{ij}=S_{i}\delta_{ij}
      +S_{i}\hat{t}^{ij}S_{j}
      +\sum_{l}S_{i}\hat{t}^{il}S_{l}\hat{t}^{lj}S_{j}
      +\dots ,
\label{Green perturb series}
\end{gather}
where $S_{i}$ is the local propagator (locator) and $\hat{t}^{ij}$ the interlayer
hopping.

Consider the term $S_{i}\hat{t}^{il}S_{l}\hat{t}^{lj}$ with $i<l<j$.
Here $t^{il}=t^{lj}= \tfrac{\Delta_{D}}{2}\tau^{+}\!\otimes\!\sigma_{0}$,
and at zero energy $S_{i}(\epsilon=0,\mathbf{k}_{\perp}=0)
=-\Delta_{S}^{-1}\tau^{x}\!\otimes\!\sigma_{0}$.  Hence
\begin{gather}
S_{i}\hat{t}^{il}S_{l}\hat{t}^{lj}
        =\frac{\Delta_{D}^{2}}{4}
          \begin{pmatrix}
            0 & 0 \\[2pt]
            0 & 1
          \end{pmatrix}
          \!\otimes\!\sigma_{0}.
\end{gather}

\begin{figure}[b!]
    \centering
    \includegraphics[width=0.8\linewidth]{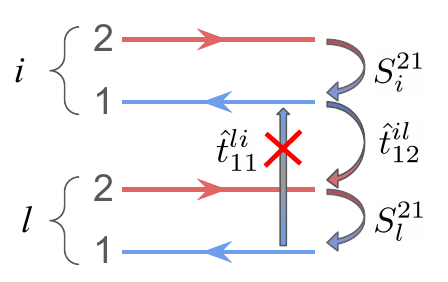}
    \caption{Graphical explanation of why any segment that \emph{returns}
    to the initial layer gives zero at $\varepsilon=0$. The diagram corresponds to the term
    $S_i \hat{t}^{il} S_l \hat{t}^{li}$ in the Green–function expansion. Propagation
    between neighbouring layers proceeds through the off–diagonal blocks
    $S^{21}_{i}$ and $S^{21}_{l}$, while the back–hopping involves
    $\hat{t}^{li}_{11}$. At zero energy the projector structure of $S_{n}$ and
    $\hat{t}$ enforces $S^{21}_{i}\hat{t}^{il}_{12}S^{21}_{l}\hat{t}^{li}_{11}=0$
    (red cross), so every diagram containing a return path vanishes. As a result,
    only strictly forward sequences survive and the zero–energy Green function
    reduces to the product shown in Eq.~(\ref{zero energy Green function}).}
    \label{fig:green_forbidden}
\end{figure}

Next consider $j=i$, which corresponds to a diagram containing one
\emph{return} to site~$i$.  Then
$t^{il}=\tfrac{\Delta_{D}}{2}\tau^{+}\!\otimes\!\sigma_{0}$ and
$t^{li}=\tfrac{\Delta_{D}}{2}\tau^{-}\!\otimes\!\sigma_{0}$, giving
\begin{gather}
S_{i}\hat{t}^{il}S_{l}\hat{t}^{li}=0 .
\end{gather}
Written with explicit indices,
$S^{21}_{i}\hat{t}^{il}_{12}S^{21}_{l}\hat{t}^{li}_{11}=0$, as illustrated
schematically in Fig.~\ref{fig:green_forbidden}.  Thus, at zero energy
every diagram containing a return segment vanishes.

Only a single, strictly forward diagram survives, so the Green function
reduces to
\begin{gather}
G_{ij}=S_{i}t^{i,i+1}S_{i+1}t^{i+1,i+2}S_{i+2}\dots
       t^{j-1,j}S_{j},
\label{zero energy Green function}
\end{gather}
which will be used below to analyze the localization length.

The localization length is obtained from the logarithm of the retarded
Green function, $\ln G^{+}(1,L)$, in the limit $L\to\infty$.  For a
matrix Green function we must consider the slowest decaying component,
so that
\begin{gather}
L_{c}^{-1}= -\lim_{L\to\infty}\frac{1}{L}\,
\Bigl\langle \ln\|G^{+}(1,L)\|\Bigr\rangle ,
\end{gather}
where $\|G^{+}(1,L)\|$ denotes the operator norm of the matrix
$G^{+}(1,L)$ (for a matrix $A$, the operator norm is
$\max\sqrt{\lambda_{i}}$, with $\lambda_{i}$ obtained from
$\det(A^{\dagger}A-\lambda I)=0$).

At zero energy the disorder Green function takes the form of
Eq.~\eqref{zero energy Green function}, which remains valid in the
presence of off‑diagonal disorder,
\begin{gather}
G(1,L)=S_{1}t^{12}S_{2}t^{23}S_{3}\ldots t^{L-1,L}S_{L}.
\end{gather}
Using
$S_{i}=-(\Delta_{S}+\eta_{i})^{-1}\tau^{x}\!\otimes\!\sigma_{0}$ and
$t^{ij}=\tfrac{\Delta_{D}^{i}}{2}\tau^{+}\!\otimes\!\sigma_{0}$ we find
\begin{gather}
G(1,L)
  =g(1,L)
   \begin{pmatrix}
     0 & 0 \\[2pt]
     1 & 0
   \end{pmatrix}\!\otimes\!\sigma_{0},
\end{gather}
with
\begin{gather}
g(1,L)=(-1)^{L}\,
       \frac{\Delta_{D}^{1}\Delta_{D}^{2}\cdots\Delta_{D}^{\,L-1}}
            {2^{L-1}\Delta_{S}^{1}\Delta_{S}^{2}\cdots\Delta_{S}^{\,L}} .
\end{gather}
Hence
\begin{gather}
\|G(1,L)\|=\bigl|g(1,L)\bigr|
\end{gather}
and the inverse localization length is
\begin{gather}
L_{c}^{-1}
  =\bigl|\langle\ln\Delta_{S}^{i}\rangle
         -\langle\ln\Delta_{D}^{i}\rangle\bigr| .
\end{gather}

We analyze three disorder scenarios:  
(i) only $\Delta_{S}$ fluctuates,  
(ii) $\Delta_{S}$ and $\Delta_{D}$ fluctuate independently,  
(iii) the two parameters fluctuate in a correlated way.  
Explicitly
\begin{enumerate}
\item $\Delta_{S}^{i}=\Delta_{S}+\eta_{i},\quad \Delta_{D}^{i}=\Delta_{D}$,
\item $\Delta_{S}^{i}=\Delta_{S}+\eta_{i},\quad \Delta_{D}^{i}=\Delta_{D}+\xi_{i}$,
\item $\Delta_{S}^{i}=\Delta_{S}+\eta_{i},\quad \Delta_{D}^{i}=\Delta_{D}-\eta_{i}$.
\end{enumerate}
For each case we employ three averaging procedures:  
(1) a uniform distribution,
$\langle\ldots\rangle_{x}=(2x_{0})^{-1}\int_{-x_{0}}^{x_{0}}\ldots\,dx$;  
(2) a Gaussian distribution,
$\langle\ldots\rangle_{x}=\sqrt{\alpha/\pi}\int_{-\infty}^{\infty}
  \ldots\,e^{-\alpha(x-\overline{x})^{2}}dx$;  
(3) a Lorentzian distribution,
$\langle\ldots\rangle_{x}=\pi^{-1}\int_{-\infty}^{\infty}
  \ldots\,\Gamma\,[\,(x-\overline{x})^{2}+\Gamma^{2}]^{-1}dx$.

All three averages can be evaluated analytically. For value $\langle\ln\left|1-\frac{x}{b}\right|\rangle$ we have:
\begin{gather}
\langle\ldots\rangle_{a}^{\text{Uniform}}=\frac{b}{2a}\ln\!\Bigl|\frac{b+a}{b-a}\Bigr|
  +\frac{1}{2}\ln\!\Bigl|1-\frac{a^{2}}{b^{2}}\Bigr|-1,\\[4pt]
\langle\ldots\rangle_{\alpha}^{\text{Gauss}}
 =-\frac{1}{2}\!\left[
   {}_{1}F_{1}^{(1,0,0)}\!\Bigl(0,\tfrac12,-\alpha b^{2}\Bigr)
   +\ln(4\alpha b^{2})+\gamma\right]\nonumber\\
 \xrightarrow[\alpha\to\infty]{}\;
   \frac{\sqrt{\pi}\,e^{-\alpha b^{2}}}{2\sqrt{\alpha b^{2}}}
   -\frac{1}{4\alpha b^{2}},\\[4pt]
\langle\ldots\rangle_{\Gamma}^{\text{Lorentz}}
 =b\!\left[1-\frac{b}{\Gamma}\arctan\!\Bigl(\tfrac{\Gamma}{b}\Bigr)\right],
\end{gather}
where ${}_{1}F_{1}^{(1,0,0)}(a,b,c)$ is the derivative of the confluent
hypergeometric function with respect to its first argument.

Substituting the boundary Green function (see
Refs.~\cite{dai2008helical,kim2015holographic,peng2017boundary}) into the
expression for $L_{c}^{-1}$ enables us to evaluate the localization
length for each distribution.  Figure~\ref{localization lenght} plots
$L_{c}$ versus disorder strength for all three cases.  For a uniform
distribution the localization length decreases as disorder grows.
Gaussian and Lorentzian disorder, which better mimic real samples, lead
to a stronger reduction and, for the Gaussian case, a divergence of
$L_{c}$.  Hence off‑diagonal disorder can enlarge the penetration depth
of edge modes, allowing them to overlap.  Such overlap enhances
inter‑mode tunneling and can compromise the topological protection; the
consequences are examined in the next part of the section.

\begin{figure}[t!]
    \centering
    \includegraphics[width=\linewidth]{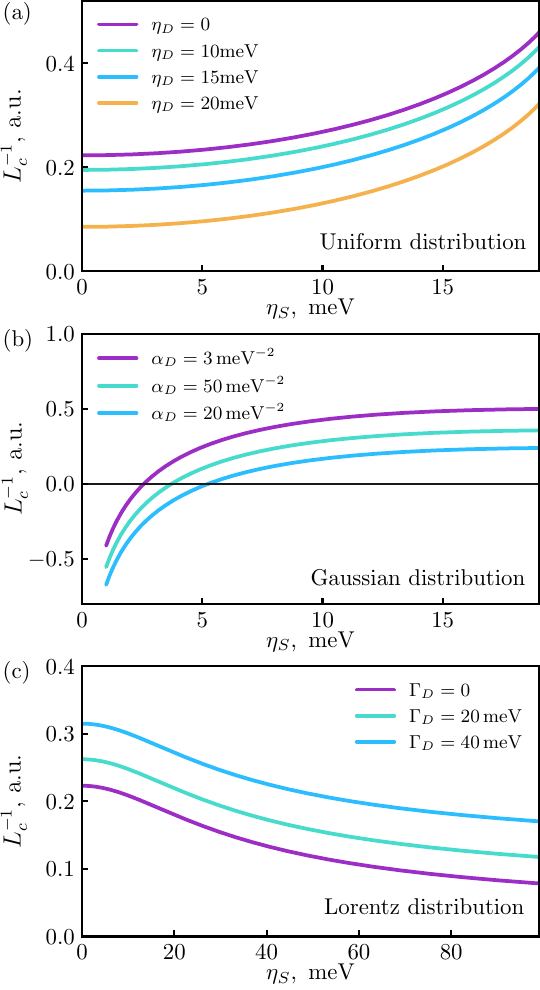}
    \caption{Inverse localization length $L_c^{-1}$ as a function of the fluctuation amplitude of the intra–layer tunneling,
    $\eta_S$, for three statistics of off–diagonal disorder.
    (a)~Uniform distribution: $L_c^{-1}$ grows monotonically with
    $\eta_S$; larger fluctuations of the inter–layer tunneling
    ($\eta_D$) systematically reduce $L_c^{-1}$.
    (b)~Gaussian distribution: $L_c^{-1}$ can change sign and approach
    zero from below as $\eta_S$ increases, signaling a disorder–driven
    delocalization ($L_c\!\to\!\infty$); the rate of this approach depends
    on the variance $\alpha_D^{-1}$ of the fluctuations of $\Delta_D$.
    (c)~Lorentzian distribution: $L_c^{-1}$ decreases with growing
    Lorentzian width $\Gamma_D$, but remains positive over the shown   interval, implying finite (though enlarged) localization length. The horizontal grey line marks $L_c^{-1}=0$.}
    \label{localization lenght}
\end{figure}

Time-reversal symmetry guarantees the dissipationless nature of edge
modes in a topological insulator, providing the familiar topological
protection.  For helical modes the absence of back‑scattering follows
from the vanishing of the matrix element
$\langle\alpha|V|\beta\rangle$--with $\alpha,\beta$ labelling
counter‑propagating modes of opposite spin--for any
${\cal T}$‑invariant potential $V$ satisfying
${\cal T}^{-1}V{\cal T}=V$.  In the ballistic limit the edge
conductance is therefore fixed by fundamental constants, as in an ideal
one‑dimensional channel.  Experimentally, however, the measured
conductance often deviates from this ideal value
\cite{konig2007quantum,konig2013spatially,olshanetsky2023observation},
implicating \emph{inelastic} scattering processes.  Several mechanisms
have been proposed.  In particular, systems with bulk inversion
asymmetry (BIA) or structural inversion asymmetry (SIA) permit inelastic
back‑scattering: while BIA/SIA preserves ${\cal T}$ symmetry (and hence
does not open a gap), it mixes states of different spin projections so
that counter‑propagating modes no longer have well‑defined spin
polarisations \cite{rothe2010fingerprint,maciejko2010magnetoconductance}.
For elastic scattering this mixing is harmless, but inelastic
processes--such as electron–phonon interactions--can then produce a finite
overlap between states of opposite momentum and degrade the edge
conductance \cite{schmidt2012inelastic}; the effect was observed
experimentally in Ref.~\cite{du2015robust}.  Additional inelastic
scattering can arise from charge puddles created by random doping
\cite{vayrynen2013helical,vayrynen2014resistance}, which also influence
thermoelectric transport in two‑dimensional topological insulators
\cite{alisultanov2025thermoelectric}.

\begin{figure}[b!]
    \centering
    \includegraphics[width=\linewidth]{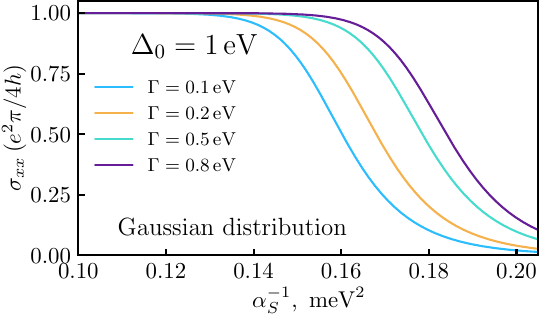}
    \caption{Edge conductivity of a three‑dimensional topological insulator versus off‑diagonal disorder in the parameter $\Delta_{S}$.}
    \label{conductivity}
\end{figure}

A further prerequisite for topological protection is that edge modes
remain confined to the sample boundaries, with no tunneling between
opposite edges.  Finite tunneling opens a gap in the edge spectrum and
permits back‑scattering.  We have shown above that off‑diagonal disorder
drives the localization length to infinity; for edge states this length
is the penetration depth.  As the depth grows the modes on opposite
edges overlap, giving a non‑zero tunneling amplitude $\Delta_t$.  Appendix~D gives
the resulting correction to the longitudinal conductance,
\begin{multline}
\sigma_{xx}\;=\;
\frac{\sigma_{xx}^{\Delta}}{\pi e^{2}/4h}
=\;
1-\frac{2}{\pi}\Bigl(1-\frac{4\Delta_{t}^{2}}{\Gamma^{2}}\Bigr)
      \\\times\arctan\!\Bigl(\frac{2\Delta_t}{\Gamma}\Bigr)
  +\frac{4\Delta_t}{\pi\Gamma}
  -\frac{4\Delta_t^{2}}{\Gamma^{2}},
\label{conductivity xx}
\end{multline}
where $\Gamma$ is the scattering rate associated with inter‑edge
tunneling.  For the tunneling amplitude we use the estimate
$\Delta_t=\Delta_{0}e^{-L/L_{c}}$, with $\Delta_{0}$ the value for fully
overlapping edges; when $L/L_{c}\gg1$, $\Delta_t\to0$.  Increasing
$L_{c}$ therefore enhances tunneling and the corresponding
conductance correction~\eqref{conductivity xx}.  Figure~\ref{conductivity}
plots $\sigma_{xx}$ versus disorder strength for a Gaussian
distribution: off‑diagonal disorder clearly degrades topological
protection, and the edge conductance is no longer fixed by fundamental
constants, providing yet another mechanism for scattering without
breaking time‑reversal symmetry.

We now turn to the gapless (Dirac) phase and ask whether
off‑diagonal disorder can localize the electronic states.  The density‑of‑states
analysis in the previous section showed that the Weyl phase is robust
even against strong disorder of this type.  A rigorous localization
study would require heavy numerics; instead we adopt Ziman’s analytic
convergence test for the Green‑function series, which, while crude,
captures the essential criterion.  A more elaborate treatment is left to
future work.

Starting from Eq.~\eqref{Green perturb series}, the local Green function
on layer~$n=0$ reads (the precise choice of layer is immaterial in the
gapless phase)
\begin{gather}
G_{00}=S_{0}
 +\sum_{j'}S_{0}t_{0j'}S_{j'}t_{j'0}S_{0}+\dots .
\label{local Green series}
\end{gather}
A state is localized on layer~$n$ if its wavefunction decays
sufficiently fast away from that layer.  Anderson’s criterion translates
this into the convergence of the series~\eqref{local Green series}.  In
contrast to Ref.~\cite{anderson1958absence}, we deal with matrix Green
functions and off‑diagonal disorder, which complicates the analysis.

Combining Eqs.~\eqref{local Green series} and
\eqref{renormalized petrurb series} we obtain the self‑energy component
\begin{gather}
\Sigma_{00}=\sum_{n}t_{0n}G_{nn}^{\bcancel{0}}t_{n0}
          =\sum_{n}t_{0n}\Lambda_{n0},
\end{gather}
with
\begin{align}
\Lambda_{n0}&=G_{nn}^{\bcancel{0}}t_{n0}
   =S_{n}t_{n0}
   +\sum_{m}S_{n}t_{nm}S_{m}t_{mn}S_{n}t_{n0} \notag\\
  &\quad +\sum_{m,p}S_{n}t_{nm}S_{m}t_{mp}S_{p}t_{pm}
           S_{m}t_{mn}S_{n}t_{n0}+\dots .
\label{self-energy part}
\end{align}

Introduce ${\cal S}_{m}=S_{n}t_{nm}$.  A generic term in the matrix
series~\eqref{self-energy part} is then a product
${\cal S}_{m}{\cal S}_{p}{\cal S}_{n}\ldots$ with an odd number of
factors and each index taking one of two values (the nearest
neighbors).  Convergence requires that every eigenvalue of each
${\cal S}_{m}$ lie inside the circle of convergence of the associated
power series, so we rewrite the expansion in terms of those eigenvalues.
Because ${\cal S}_{m}$ is not necessarily Hermitian, we introduce right
and left eigenvectors
\begin{gather}
{\cal S}_{m}\,|\psi_{m}^{R}\rangle=\lambda_{m}\,|\psi_{m}^{R}\rangle,\\
{\cal S}_{m}^{\dagger}\,|\psi_{m}^{L}\rangle=\lambda_{m}\,|\psi_{m}^{L}\rangle,
\end{gather}
with the bi‑orthogonality condition
$\langle\psi_{n}^{L}|\psi_{m}^{R}\rangle=\delta_{mn}$.  When
${\cal S}_{m}$ is Hermitian the left and right eigenvectors coincide.
In that case
\begin{equation}
\lambda_{m}^{1,2}=
\frac{1}{2}\,
\frac{\Delta_{D}\bigl(\Delta_{S}+\eta_{S}^{n}\bigr)}
     {\varepsilon_{\perp}^{2}-\bigl(\Delta_{S}+\eta_{S}^{n}\bigr)^{2}}\,
     \bigl(\delta_{m,n-1},\delta_{m,n+1}\bigr),
\label{eigenvalues Ziman}
\end{equation}
where $\varepsilon_{\perp}^{2}=\varepsilon^{2}-v_{F}^{2}k_{\perp}^{2}$.
The in‑plane helical term
$\tau^{z}v_{F}(\hat{\bm{z}}\!\times\!\bm{\sigma})\!\cdot\!\mathbf{k}_{\perp}$
enters only through $\varepsilon_{\perp}$ and does not change the
structure of $\lambda_{m}$.  Since adjacent indices in the product
${\cal S}_{m}{\cal S}_{p}{\cal S}_{n}\ldots$ differ by one, we may
replace each ${\cal S}_{n}$ by
$\lambda_{n}=\frac{1}{2}\Delta_{D}(\Delta_{S}+\eta_{S}^{n})
             /[\varepsilon_{\perp}^{2}-(\Delta_{S}+\eta_{S}^{n})^{2}]$,
so that
\begin{equation}
\Lambda_{n0}\;\to\;
\lambda_{0}
+\sum_{m}\lambda_{m}\lambda_{n}\lambda_{0}
+\sum_{m,p}\lambda_{m}\lambda_{p}\lambda_{m}\lambda_{n}\lambda_{0}+\ldots .
\end{equation}

We now apply Ziman’s approach.  His key step is to replace products such
as ${\lambda}_{i}\lambda_{i'}{\lambda}_{i''}\dots$ by
\begin{gather}
{\lambda}_{i}\lambda_{i'}{\lambda}_{i''}\dots
\;\Longrightarrow\;
c\,\exp\!\Bigl\langle
        \ln\bigl|{\lambda}_{i}\lambda_{i'}{\lambda}_{i''}\dots\bigr|
       \Bigr\rangle ,
\end{gather}
where $\langle\cdots\rangle$ denotes disorder averaging and the constant
$c$ counts nearest neighbors.  Each layer has two neighbors, so
$c=2^{L}$ for a product of $L$ factors.  Hence a typical term becomes
\[
2^{L}\exp\!\Bigl\langle\ln|{\lambda}_{i}|\Bigr\rangle^{\,L}
     =\bigl(2\,\exp\langle\ln|{\lambda}_{i}|\rangle\bigr)^{L},
\]
and the series converges absolutely provided
$\exp\langle\ln|2{\lambda}_{i}|\rangle<1$.

Consider the energy $\varepsilon=v_{F}k_{\perp}$--the Dirac (Weyl) point
in the gapless phase--and set $\Delta_{S}=\Delta_{D}$.  Then
\[
\lambda_{n}\bigl(\varepsilon=v_{F}k_{\perp}\bigr)
     =-\frac12\frac{\Delta_{S}}{\Delta_{S}+\eta_{S}^{n}},
\]
so that
\begin{gather}
\Bigl\langle\ln|2{\lambda}_{i}|\Bigr\rangle
=1-\frac12\Bigl[\,
      \ln\bigl|1-\delta_{0}^{2}\bigr|
     +\frac{1}{\delta_{0}}
      \ln\bigl|\tfrac{1+\delta_{0}}{1-\delta_{0}}\bigr|
     \Bigr],
\end{gather}
with $\delta_{0}=\eta_{0}/\Delta_{S}$.  One checks immediately that the
convergence condition fails even for $\delta_{0}=1$.  Hence, in the
Dirac semimetal regime, off‑diagonal disorder does not localize the
states and cannot open a gap.

\section{Conclusion}
The present work delivers a systematic analysis of multilayer topological insulators subject to off‑diagonal disorder, i.e.\ random variations of the interlayer tunneling amplitudes that couple surface states belonging to adjacent layers. Two independent self‑consistent Green‑function approaches were developed: the locator expansion in localized states and a $t$‑matrix\,+\,RPA expansion in Bloch states. Both give mutually consistent results and confirm that a single Hermitian tunneling defect always generates an in‑gap bound state whose energy crosses the mid‑gap level only in the trivial phase, never in the topological phase. Adding non‑Hermitian terms splits this level but preserves the same crossing rule, so the behavior of the bound state remains a local diagnostic of bulk topology.

When such defects are present with finite density, off‑diagonal disorder fills the gap with bulk states and can eventually close it. The Weyl semimetal phase survives large fluctuations, whereas the anomalous quantum Hall phase is fragile and its Hall plateau shrinks as disorder increases. Off‑diagonal disorder also lengthens the penetration depth of edge modes; for a uniform distribution this effect is modest, but Gaussian or Lorentzian disorder can drive a divergence of the localization length, so that modes on opposite edges overlap and acquire a finite tunneling amplitude. The resulting scattering rate produces a correction to the longitudinal conductance, reducing it below the quantized value without breaking time‑reversal symmetry.

Finally, an analytic convergence test shows that in the gapless Dirac (or Weyl) phase the Green‑function series never converges, indicating that off‑diagonal disorder cannot localize the zero‑energy states or open a gap. Taken together, these results explain how random interlayer hopping can erode the ideal transport signatures expected for multilayer topological insulators and suggest experimental observables--such as a narrowed Hall plateau or a disorder-induced suppression of edge conductance--that can be used to diagnose the presence and strength of off-diagonal disorder in real samples.

Looking forward, it will be important to extend the present analysis to include electron–electron interactions and realistic multilayer geometries, to investigate possible disorder‑driven topological superconducting phases, and to compare the predicted localization–delocalization crossover with large‑scale numerical simulations and forthcoming transport measurements on magnetic van der Waals heterostructures.

\begin{acknowledgments}
The study is supported by the Ministry of Science and Higher Education of the Russian Federation (Goszadaniye), project No. FSMG-2023-0011.  The work of A.K. is supported by the Icelandic Research Fund (Ranns\'oknasj\'o{\dh}ur, Grant No.~2410550). 
\end{acknowledgments}

\appendix

\section{Derivation of perturbation series for matrix Green's function}

We have Hamiltonian (here we have rewritten Hamiltonian~\eqref{off-diagonal Hamiltonian} in tensor index form)
\begin{gather}
\mathcal{H}=\sum_{\mathbf{k}_{\perp},i}c_{\mathbf{k}_{\perp}i}^{\dagger}\otimes\left[ h\left(\mathbf{k}_{\perp}\right)c_{\mathbf{k}_{\perp}i}+t^{ij}c_{\mathbf{k}_{\perp}j}\right]=\nonumber\\
=\sum_{\mathbf{k}_{\perp},i}c_{\mathbf{k}_{\perp}i}^{\alpha\dagger}h_{\alpha\beta}c_{\mathbf{k}_{\perp}i}^{\beta}+\sum_{\mathbf{k}_{\perp},i,j}c_{\mathbf{k}_{\perp}i}^{\alpha\dagger}t^{ij}_{\alpha\beta}c^{\beta}_{\mathbf{k}_{\perp}j}\label{off-diagonal Hamiltonian_indexes},
\end{gather}
where $c_{\mathbf{k}_{\perp}i}=\left(c_{\mathbf{k}_{\perp}i}^{1}~c_{\mathbf{k}_{\perp}i}^{2}\right)^T$ are annihilation operators in subspace that is related with the edge modes (it is denoted as $\tau$) of topological layer (each such operator is a spinor with respect to the spin degrees of freedom (it is denoted as $\sigma$): $c_{\mathbf{k}_{\perp}i}^{\alpha}=\left(c_{\mathbf{k}_{\perp}i}^{\alpha\uparrow}~c_{\mathbf{k}_{\perp}i}^{\alpha\downarrow}\right)$), $h\left(\mathbf{k}_{\perp}\right)=\upsilon_{F}\tau^{z}\otimes\left(\hat{\bm{z}}\times\bm{\sigma}\right)\cdot\mathbf{k}_{\perp}+\Delta_{S}\tau^{x}\otimes\sigma_0$ and $t^{ij}=\frac{\Delta_{D}}{2}\left[\tau^{+}\otimes\sigma_0\delta_{j,i+1}+\tau^{-}\otimes\sigma_0\delta_{j,i-1}\right]$, indexes $\alpha,\beta$ denote upper and lower edge modes basis. For operators $c,c^{\dagger}$ we have 
\begin{gather}
c_{\mathbf{k}i}^{\alpha}c_{\mathbf{k'}j}^{\beta\dagger}+c_{\mathbf{k'}j}^{\beta\dagger}c_{\mathbf{k}i}^{\alpha}=\delta_{\alpha\beta}\delta_{ij}\delta_{\mathbf{k'}\mathbf{k}}.
\end{gather}
We introduce the Green's function as follows
\begin{gather}
G_{ij}=\langle c_{\mathbf{k}_{\perp}i}\otimes c_{\mathbf{k}_{\perp}j}^{\dagger}\rangle=\nonumber\\
=\begin{pmatrix}
    \langle c_{\mathbf{k}_{\perp}i}^{1}\otimes c_{\mathbf{k}_{\perp}j}^{1\dagger}\rangle&\langle c_{\mathbf{k}_{\perp}i}^{1}\otimes c_{\mathbf{k}_{\perp}j}^{2\dagger}\rangle\\
    \langle c_{\mathbf{k}_{\perp}i}^{2}\otimes c_{\mathbf{k}_{\perp}j}^{1\dagger}\rangle&\langle c_{\mathbf{k}_{\perp}i}^{2}\otimes c_{\mathbf{k}_{\perp}j}^{2\dagger}\rangle
\end{pmatrix}.
\end{gather}
We use the following equation for Green's function
\begin{equation}
\epsilon\langle A~B \rangle=\langle \left[A,B\right] \rangle+\langle \left[A,H\right]~B \rangle.
\end{equation}
where $A,B$ are some operators. Then we have
\begin{gather}
\epsilon G_{ij}=\delta_{ij}I_{4\times4}+\langle\left(\hat{h} c_{\mathbf{k_{\perp}}i}\right)\otimes c_{\mathbf{k_{\perp}}j}^{\dagger}\rangle+\nonumber\\
+\sum_{l}\langle\left(\hat{t}^{il} c_{\mathbf{k_{\perp}}l}\right)\otimes c_{\mathbf{k_{\perp}}j}^{\dagger}\rangle.
\end{gather}
Using the following property of matrices $(AB)\otimes B^T=A (B\otimes B^T)$, where $A$ is matrix and $B$ is vector, we have
\begin{gather}
\left(\epsilon-\hat{h}\right) G_{ij}=\delta_{ij}+
\sum_{l}\hat{t}^{il}G_{lj}.
\end{gather}
Now we can introduce zero local Green's function $S_i=\left(\epsilon-\hat{h}\right)^{-1}$ and put it to right part. Then
\begin{gather}
G_{ij}=S_{i}\delta_{ij}+S_{i}\hat{t}^{ij}S_{j}+\sum_{l}S_{i}\hat{t}^{il}S_{l}\hat{t}^{lj}S_{j}+\dots .\label{Green perturb series}
\end{gather}
From this equation we obtain:
\begin{gather}
G_{i\neq j}=\sum_{l}S_{i}\hat{t}^{il}S_{l}\hat{t}^{lj}S_{j}+\dots . 
\end{gather}

\section{Expression for local Green function}

We now compute $G_{ll}^{0}$ explicitly in the absence of Zeeman splitting.  
(The integral for finite splitting is performed analogously, but the resulting expressions are unwieldy.)
\begin{gather}
G_{ll}^{0}=\sum_{k_{z}}G_{\mathbf{k}}^{0}\to
\int_{-\pi}^{\pi}\frac{dk_{z}}{2\pi}\,
\frac{\varepsilon+\mathcal{H}_{\mathbf{k}}}{\varepsilon^{2}-\upsilon_{F}^{2}k_{\perp}^{2}-\Delta^{2}(k_{z})}
=\frac{1}{4\pi i}\oint dz\nonumber\\
\times\frac{\Delta_{D}\tau^{x}(z^{2}+1)+2\bigl(\varepsilon+\tau^{z}\upsilon_{F}(\hat{\bm{z}}\!\times\!\bm{\sigma})\!\cdot\!\mathbf{k}_{\perp}+\Delta_{S}\tau^{x}\bigr)z}{\bigl(\Delta_{S}\Delta_{D}z^{2}-(\varepsilon^{2}-\upsilon_{F}^{2}k_{\perp}^{2}-\Delta_{S}^{2}-\Delta_{D}^{2})z+\Delta_{S}\Delta_{D}\bigr)z}\!,
\end{gather}
where the contour runs along the unit circle.  
The integrand has poles at $z=0$ and at $z_{1,2}$, with
\begin{gather}
z_{1,2}=\frac{\varepsilon_{\perp}^{2}-\Delta_{S}^{2}-\Delta_{D}^{2}}{2\Delta_{S}\Delta_{D}}
\pm\Bigl|\sqrt{\frac{(\varepsilon_{\perp}^{2}-\Delta_{S}^{2}-\Delta_{D}^{2})^{2}}{4\Delta_{S}^{2}\Delta_{D}^{2}}-1}\Bigr|,
\end{gather}
and $\varepsilon_{\perp}^{2}=\varepsilon^{2}-\upsilon_{F}^{2}k_{\perp}^{2}$.  
Energies satisfying $\!(\Delta_{S}\!-\!\Delta_{D})^{2}<\varepsilon_{\perp}^{2}<(\Delta_{S}\!+\!\Delta_{D})^{2}$ lie in the allowed bands; the corresponding roots are complex and satisfy $\lvert z_{1,2}\rvert=1$.  
For $\varepsilon_{\perp}^{2}<(\Delta_{S}-\Delta_{D})^{2}$ and $\varepsilon_{\perp}^{2}>(\Delta_{S}+\Delta_{D})^{2}$ the roots become real; this interval therefore represents the forbidden region.  
Only the root with the minus sign (since $x-\sqrt{x^{2}-1}<1$ for $x>1$) falls inside the contour, and the residue theorem yields
\begin{equation}
G_{ll}^{0}=-\frac{\bigl(z_{1}\Delta_{D}+\Delta_{S}\bigr)\tau^{x}
              +\varepsilon
              +\tau^{z}\upsilon_{F}(\hat{\bm{z}}\!\times\!\bm{\sigma})\!\cdot\!\mathbf{k}_{\perp}}
             {2\Delta_{S}\Delta_{D}
              \sqrt{\frac{(\varepsilon_{\perp}^{2}-\Delta_{S}^{2}-\Delta_{D}^{2})^{2}}{4\Delta_{S}^{2}\Delta_{D}^{2}}-1}},
\label{Green_ll}
\end{equation}
valid for $\varepsilon_{\perp}^{2}<(\Delta_{S}-\Delta_{D})^{2}$, and
\begin{gather}
G_{ll}^{0}=\frac{\tau^{x}\bigl(z_{2}\Delta_{D}+\Delta_{S}\bigr)
               +\varepsilon
               +\tau^{z}\upsilon_{F}(\hat{\bm{z}}\!\times\!\bm{\sigma})\!\cdot\!\mathbf{k}_{\perp}}
              {2\Delta_{S}\Delta_{D}
               \sqrt{\frac{(\varepsilon_{\perp}^{2}-\Delta_{S}^{2}-\Delta_{D}^{2})^{2}}{4\Delta_{S}^{2}\Delta_{D}^{2}}-1}}
\label{Green_ll+}
\end{gather}
for $\varepsilon_{\perp}^{2}>(\Delta_{S}+\Delta_{D})^{2}$.  
In both cases $G_{ll}^{0}$ is purely real, so the DOS--proportional to its imaginary part--vanishes, consistent with a gap.  
Note that in the limits $\Delta_{D}\to0$ and $\mathbf{k}_{\perp}\to0$, Eqs.~\eqref{Green_ll} and \eqref{Green_ll+} reduce to the familiar one‑dimensional lattice result.

For $\varepsilon_{\perp}=0$ one finds
\begin{align}
z_{1,2}=-\frac{\Delta_{S}^{2}+\Delta_{D}^{2}\mp\lvert\Delta_{S}^{2}-\Delta_{D}^{2}\rvert}{2\Delta_{S}\Delta_{D}}.
\end{align}
Thus, in the trivial insulator ($\Delta_{S}>\Delta_{D}$) one has $z_{1}=\Delta_{D}/\Delta_{S}$, while in the topological insulator ($\Delta_{S}<\Delta_{D}$) $z_{1}=\Delta_{S}/\Delta_{D}$.  
At the level of the Green function, no qualitative distinction arises between the two phases.

For the off‑diagonal Green function we set  
$G_{lm}^{0}=G_{ll}^{0}\,e^{-a(\varepsilon,\mathbf{k}_{\perp})|l-m|}$, where
\begin{gather}
a=\frac{1}{\ln\Bigl|\dfrac{\varepsilon_{\perp}^{2}-\Delta_{S}^{2}-\Delta_{D}^{2}}{2\Delta_{S}\Delta_{D}}
      -\Bigl|\sqrt{\dfrac{(\varepsilon_{\perp}^{2}-\Delta_{S}^{2}-\Delta_{D}^{2})^{2}}
                         {4\Delta_{S}^{2}\Delta_{D}^{2}}-1}\Bigr|\Bigr|}.
\end{gather}
The quantity $a$ is finite and positive whenever $\Delta_{S}\neq\Delta_{D}$, signaling localization; it vanishes at $\Delta_{S}=\Delta_{D}$, consistent with the delocalized Dirac semimetal.

For energies inside the allowed bands we analytically continue Eq.~\eqref{Green_ll} to the interval $(\Delta_{S}-\Delta_{D})^{2}<\varepsilon_{\perp}^{2}<(\Delta_{S}+\Delta_{D})^{2}$, obtaining
\begin{gather}
\text{Re}\,G_{ll}^{0\pm}=-\frac{\tau^{x}}{\Delta_{S}},\\
\text{Im}\,G_{ll}^{0\pm}=\mp
\frac{
  \dfrac{\varepsilon_{\perp}^{2}+\Delta_{S}^{2}-\Delta_{D}^{2}}{2\Delta_{S}}\tau^{x}
  +\varepsilon
  +\tau^{z}\upsilon_{F}(\hat{\bm{z}}\!\times\!\bm{\sigma})\!\cdot\!\mathbf{k}_{\perp}}
{\Delta_{S}\Delta_{D}\sqrt{1-\dfrac{(\varepsilon_{\perp}^{2}-\Delta_{S}^{2}-\Delta_{D}^{2})^{2}}
                                      {4\Delta_{S}^{2}\Delta_{D}^{2}}}}.
\end{gather}
The density of states then follows from  
\begin{gather}
\rho_{0}(\varepsilon)=\mp\pi^{-1}\sum_{\mathbf{k}_{\perp}}
\operatorname{tr}\bigl(\text{Im}\,G_{ll}^{0\pm}\bigr).
\end{gather}

\section{Approximate solution of the equation for the distribution function $f\left(w\right)$ for zero energy}

First, we recast the expansion in Eq.~\eqref{Green perturb series} within the framework of \textit{renormalized} perturbation theory (see, e.g., Ref.~\cite{economou2006green}):
\begin{gather}
G_{nn}=S_{n}
+\sum_{m=n\pm1}G_{nn}\,t_{n,m}\,G_{m}^{\bcancel{n}}\,t_{m,n}\,S_{n},
\label{renormalized petrurb series}
\end{gather}
where $t_{ij}\neq0$ only for nearest neighbors $j=i\pm1$, and $G_{m}^{\bcancel{n}}$ denotes the Green function for all trajectories that start and end on layer $m\neq n$ without crossing layer~$n$.  
The quantity $G_{n+1}^{\bcancel{n}}$ obeys
\begin{gather}
G_{n+1}^{\bcancel{n}}
=S_{n+1}
+G_{n+1}^{\bcancel{n}}\,t_{n+1,n+2}\,
 G_{n+2}^{\bcancel{n+1}}\,t_{n+2,n+1}\,S_{n+1}.
\end{gather}
Note that the notation $\bcancel{n}$ is equivalent to taking the limit $V_{n}\!\to\!\infty$; in that limit $S_{n}\!\to\!0$, so paths that cross layer~$n$ are excluded.  Rewriting the last equation gives
\begin{gather}
G_{n+1}^{\bcancel{n}}
=\Bigl(S_{n+1}^{-1}
  -\tfrac{\Delta_{D}^{2}}{4}\,
   \tau^{+}\,G_{n+2,n+2}^{\bcancel{n+1}}\tau^{-}\Bigr)^{-1},
\end{gather}
where we used $t_{n,n+1}=\tfrac{\Delta_{D}}{2}\tau^{+}$ and
$t_{n+1,n}=\tfrac{\Delta_{D}}{2}\tau^{-}$.

Crucially, $G_{n+2,n+2}^{\bcancel{n+1}}$ is independent of the local
potentials $V_{n+1},V_{n},\dots$ because $V_{n+1}\!\to\!\infty$ removes layer
$n+1$ from the relevant trajectories; only layers $n+2,n+3,\dots$ enter
this Green function.  This observation will be essential below.

The quantities
$S_{n+1}^{-1}=\varepsilon-(\Delta_{S}+\eta_{n+1})\tau^{x}$,
$G_{n+1}^{\bcancel{n}}$, and $G_{n+2}^{\bcancel{n+1}}$ are therefore
random matrices linked through the stochastic variable
$\eta_{n+1}$.  Let $P(\eta_{n})$ be the probability distribution of
$\eta_{n}$, and denote
$w = G_{n+1}^{\bcancel{n}}$, $w' = G_{n+2}^{\bcancel{n+1}}$.  
Define $f(\varepsilon,w)$ and $f(\varepsilon,w')$ as the corresponding
probability distributions of the matrix elements of $w$ and $w'$.
Because $w'$ does not depend on $\eta_{n+1}$, the random variables
$w'$ and $\eta_{n+1}$ are statistically independent.

Next we derive an equation that relates the distributions
$f\!\left(\varepsilon,w\right)$, $f\!\left(\varepsilon,w'\right)$ and
$P\!\left(V_{m}\right)$.  The task is complicated by the fact that $w$
and $w'$ are \emph{random matrices} rather than $c$‑numbers.  For the
joint probability density $f\!\left(w,w'\right)$ one may write
\begin{equation}
f_{w,w'}
  =P\!\left(\varepsilon-\frac{\Delta_{D}^{2}}{4}\tau^{+}w'\tau^{-}-w^{-1}\right)
   f\!\left(w'\right)\,
   \bigl|J_{w\to V_{m}}\bigr|,
\end{equation}
where $J_{w\to V_{m}}$ is the matrix Jacobian of the transformation
$w\!\to V_{m}$.  For matrices the Jacobian is expressed through the
Fréchet derivative,
\begin{gather}
\bigl|J_{w\to V_{m}}\bigr|
  =\Bigl|\det
     \Bigl(\frac{\partial\,\mathrm{vec}(V_{m})}
                {\partial\,\mathrm{vec}(w)}\Bigr)
    \Bigr|,
\end{gather}
with $\mathrm{vec}$ the standard vectorisation map.  In the present
case
\begin{gather}
\frac{\partial\,\mathrm{vec}
        \bigl(\varepsilon-\tfrac{\Delta_{D}^{2}}{4}\tau^{+}w'\tau^{-}-w^{-1}\bigr)}
     {\partial\,\mathrm{vec}(w)}
  =w^{-2}\!\otimes I_{4\times4},
\end{gather}
and therefore
\begin{gather}
\bigl|J_{w\to V_{m}}\bigr|
  =\bigl|\det w\bigr|^{-8}.
\end{gather}
Integrating the joint density over $w'$ yields
\begin{multline}
f\!\left(w\right)
  =\bigl|\det w\bigr|^{-8}\\
\times\int
   P\!\left(\varepsilon
            -\frac{\Delta_{D}^{2}}{4}\tau^{+}w'\tau^{-}-w^{-1}\right)
   f\!\left(w'\right)\,dw',
\label{distribution function}
\end{multline}
where $dw'$ is the natural measure on the manifold of matrices $w'$.
The Green function $G\!\left(1,L\right)$ can be written as a series
analogous to Eq.~\eqref{local Green series} and renormalized with
respect to~$V_{n}$:
\begin{gather}
G\!\left(1,L\right)
  =G_{11}\,t_{12}\,G_{22}^{\bcancel{1}}\,t_{23}\,
   G_{33}^{\bcancel{2}}\ldots
   G_{L-1,L-1}^{\bcancel{L-2}}\,t_{L-1,L}\,G_{L}^{0},
\end{gather}
with $t_{12}=t_{23}=\ldots=t_{L-1,L}
  =\tfrac{\Delta_{D}}{2}\tau^{+}$.  
In general the operator norm of a product does not factor into the
product of norms unless the matrices commute, which they do not here;
nevertheless one can use
$\|AB\cdots\|\le\|A\|\,\|B\|\cdots$.  
Because we are interested in the possibility that the localization
length diverges, i.e.\ $L_{c}^{-1}\!\to\!0$, it is sufficient--and
rigorous for that limit--to approximate
$\|AB\cdots\|\approx\|A\|\,\|B\|\cdots$.  If localization length
diverges under this approximation it must also diverge in the exact
treatment.  Hence,
\begin{multline}
L_{c}^{-1}\approx
-\frac{1}{2}\,\Delta_{D}\,
 \Bigl\langle
   \ln\bigl\|G_{\,n+1,n+1}^{\bcancel{n}}\,\tau^{+}\bigr\|
 \Bigr\rangle
\\
=-\frac{1}{2}\,\Delta_{D}\,
  \int f(w)\,\ln\bigl\|w\,\tau^{+}\bigr\|\,dw,
\end{multline}
where the explicit layer index appears only inside the disorder
average.  Finally, employing Eq.~\eqref{distribution function} gives the
desired expression
\begin{gather}
L_{c}^{-1}
  =-\frac{1}{2}\,\Delta_{D}\,
    \int f(w)\,\ln\|w\|\,dw .
\end{gather}

\section{Correction to the $\sigma_{xx}$ due to modes tunneling. Simplest approach}

When tunneling between edge modes is introduced, the helical protection is partially lifted. Here we compute $\sigma_{xx}$. 

We consider two coupled Dirac surfaces (top and bottom) with tunneling amplitude $\Delta\sim e^{-L/L_c}$: 
\begin{gather}
H=\upsilon_F\left(\sigma_x k_y-\sigma_y k_x\right)\tau_z+\Delta\tau_x,
\end{gather}
Next, we diagonalize $H$ by transforming to the chiral basis:
\begin{gather}
H=
\begin{pmatrix}
    \Delta&\upsilon_F\left(k_y+ik_x\right)\\
    \upsilon_F\left(k_y-ik_x\right)&-\Delta
\end{pmatrix}
\end{gather}
This Hamiltonian has the following eigenvalues and eigenvectors
\begin{gather}
E_{\pm}\left(k\right)=\pm\sqrt{\upsilon_F^2 k^2+\Delta^2},\\
\left|+\right\rangle=\begin{pmatrix}
    \cos\frac{\theta}{2}\\
    \sin\frac{\theta}{2}e^{i\phi_k}
\end{pmatrix},~
\left|-\right\rangle=\begin{pmatrix}
    -\sin\frac{\theta}{2}\\
    \cos\frac{\theta}{2}e^{i\phi_k}
\end{pmatrix},
\end{gather}
where $\tan\theta=\upsilon_F k/\Delta,~\phi_k=\tan^{-1}\left(k_y/k_x\right)$. These eigenvectors are orthonormal. Let's find matrix elements of the velocity operator $v_x=\partial H/\partial k_x$. We have 
\begin{gather}
\left\langle+\right|v_x\left|+\right\rangle=\upsilon_F \sin\theta\cos\phi_k,\\
\left\langle-\right|v_x\left|-\right\rangle=\upsilon_F \sin\theta\sin\phi_k,\\
\left\langle\pm\right|v_x\left|\mp\right\rangle=\upsilon_F\left(\pm i\cos\phi_k-\cos\theta\sin\phi_k\right)
\end{gather}
To calculate the DC conductivity $\sigma_{xx}$, we used the Kubo formula~\cite{shon1998quantum,nomura2007quantum}
\begin{multline}
\sigma_{xx}=e^2\hbar\int \frac{d^2k}{2\pi^2}\\
\times\sum_{m,n}\frac{f\left(E_n\right)-f\left(E_m\right)}{E_m-E_n}\frac{\left|\left\langle n\right|v_x\left| m\right\rangle\right|^2 \Gamma}{\left(E_n-E_m\right)^2+\Gamma^2}.
\end{multline}
In general, both intra and inter band terms give contributions to the conductivity. Importantly, that in undoped case ($E_F=0$) the interband term does not vanishes for gapless spectrum and gives conductivity's minimum $\approx e^2/h$ (quantum correction). Let's show it for ideal case without bandgap.

In the ideal case (there is time reversal symmetry and no tunneling between different edge modes) we have $\Gamma\to 0$. Time-reversal symmetry enforces orthogonality between counter-propagating states - Kramer's theorem. This makes the backscattering rate $\Gamma\sim |\left\langle+\right|V\left|-\right\rangle|^2=0$ for any TRS-preserving perturbation $V$. In this limit we have
\begin{gather}
\lim_{\Gamma\to0}\frac{\Gamma}{\left(E_n-E_m\right)^2+\Gamma^2}\to\pi\delta\left(E_n-E_m\right).
\end{gather}
Then for intraband contribution we have
\begin{gather}
\sigma_{xx}^{intra}=2e^2\tau\int_{0}^{\infty} EdE \left(-\frac{\partial f}{\partial E}\right) \xrightarrow{\mu\to0} 0.
\end{gather}
where $\tau$ is an elastic and small-angle scattering of electrons on the surface of topological insulator. This contribution is Drude-like (classical) and vanishes for the undoped case ($E_F=0$). The inter band contribution is more interesting for gapless case. In this case we have
\begin{multline}
\sigma_{xx}^{inter}=2e^2\hbar\int \frac{d^2k}{2\pi^2}\\
\times\frac{f\left(E_+\right)-f\left(E_-\right)}{E_{-}-E_{+}}\left|\left\langle +\right|v_x\left| -\right\rangle\right|^2
\pi\delta\left(E_+-E_-\right).
\end{multline}
For undoped case we have: $f\left(E_+\right)\to0,~f\left(E_-\right)\approx1$. Taking into account that $E_{-}-E_{+}=-2\upsilon_F\hbar k$ and $d^2k=d\phi_k kdk$, we obtain
\begin{gather}
\sigma_{xx}^{0}=\frac{\pi e^2}{4h}.
\end{gather}

Let's now investigate the case of non-zero tunneling when $\Delta\neq0$. At presence of tunneling we can put $\Gamma\propto\Delta$. We will use the expression: $\Gamma=\gamma\Delta+\hbar\tau^{-1}$. In this case the intraband part will contains exponential factor $\exp\left(-\Delta/T\right)$ at $E_F=0$. Let's calculate the interband part of conductivity. In this case we have
\begin{gather}
\int_0^{2\pi}d\phi_k|\left\langle +\right|v_x\left|-\right\rangle|^2=\pi\upsilon_F^2\left(1+\frac{\Delta^2}{E^2}\right)
\end{gather}
Then
\begin{gather}
\sigma_{xx}^{\Delta}=\frac{e^2}{h}\int_{\Delta}^{\infty} dx
\left(1+\frac{\Delta^2}{x^2}\right)
\frac{\Gamma}{4x^2+\Gamma^2}.
\end{gather}
After integration we finally obtain
\begin{multline}
\delta\sigma_{xx}=\frac{\sigma_{xx}^{\Delta}-\sigma_{xx}^0}{\pi e^2/4h}\\
=-\frac{2}{\pi}\left(1-\frac{4\Delta^2}{\Gamma^2}\right)\arctan\frac{2\Delta}{\Gamma}+\frac{4\Delta}{\pi\Gamma}-\frac{4\Delta^2}{\Gamma^2}.
\end{multline}

\bibliography{apssamp}
\end{document}